\documentclass[fleqn,usenatbib]{mnras}

\usepackage{newtxtext,newtxmath}
\usepackage[T1]{fontenc}

\usepackage{caption}

\DeclareRobustCommand{\VAN}[3]{#2}
\let\VANthebibliography\thebibliography
\def\thebibliography{\DeclareRobustCommand{\VAN}[3]{##3}\VANthebibliography}

\usepackage{graphicx}	% Including figure files
\usepackage{amsmath}	% Advanced maths commands
\usepackage{amssymb}	% Extra maths symbols

\title[The Megamaser Cosmology Project. XII. ]{The Megamaser Cosmology Project. XII. \\ 
VLBI Imaging of H$_{2}$O Maser Emission in Three Active Galaxies
and the Effect of AGN Winds on Disk Dynamics}

\author[C. Y. Kuo et al.]{
C. Y. Kuo,$^{1,2}$\thanks{E-mail: cykuo@mail.nsysu.edu.tw (NSYSU)}
J. A. Braatz,$^{3}$
C. M. V. Impellizzeri,$^{3,4}$
F. Gao,$^{5}$
D. Pesce,$^{6,7}$
M. J. Reid,$^{6}$  \newauthor
J. Condon,$^{3}$
F. Kamali,$^{8}$
C. Henkel,$^{8,9}$ 
J. E. Greene,$^{10}$
\\
$^{1}$ Physics Department, National Sun Yat-Sen
University, No. 70, Lien-Hai Rd, Kaosiung City 80424, Taiwan, R.O.C\\
$^{2}$Academia Sinica Institute of Astronomy and Astrophysics, P.O. Box 23-141, Taipei 10617, Taiwan, R.O.C.\\
$^{3}$ National Radio Astronomy Observatory, 520 
Edgemont Road, Charlottesville, VA 22903, USA\\
$^{4}$ Joint Alma Office, Alsonso de Cordova 3107,
Vitacura, Santiago, Chile\\
$^{5}$ Max Planck Institute for extraterrestrial
Physics, Giessenbach str., 85748, Garching, Germany\\
$^{6}$ Center for Astrophysics $|$ Harvard \& Smithsonian,
60 Garden Street, Cambridge, MA 02138, USA \\
$^{7}$ Black Hole Initiative at Harvard University,
20 Garden Street, Cambridge, MA 02138, USA \\
$^{8}$ Max-Planck-Institut f\"ur Radioastronomie, 
Auf dem H\"ugel 69, 53121 Bonn, Germany\\
$^{9}$ Astron. Dept., King Abdulaziz University,
PO Box 80203, 21589 Jeddah, Saudi Arabia\\
$^{10}$Department of Astrophysical Sciences, Princeton 
University, Princeton, NJ 08544, USA
}

\begin{document}
\label{firstpage}
\pagerange{\pageref{firstpage}--\pageref{lastpage}}
\maketitle

\begin{abstract}
We present VLBI images and kinematics
of water maser emission in three active galaxies: NGC 5728, Mrk
1, and IRAS 08452$-$0011.  IRAS 08452$-$0011, at a distance of $\sim$200 Mpc, is a triple-peaked H$_{2}$O
megamaser, consistent with a Keplerian rotating disk, indicating
a black hole mass of (3.3$\pm$0.2)$\times$10$^{7}$ $M_{\odot}$.
NGC 5728 and Mrk 1 display double-peaked spectra and 
VLBI imaging reveal complicated gas kinematics that
do not allow for a robust determination of black hole mass.
We show evidence that the masers in NGC 5728 are in a wind while the Mrk 1 maser  system has both disk and outflow components. We also find that disturbed morphology and kinematics
are a ubiquitous feature of all double-peaked maser systems, implying
that these maser sources may reside in environments where AGN winds
are prominent at $\sim$1 pc scale and have significant impact on
the masing gas.  Such AGN tend to have black hole masses 
$M_{\rm BH}$ $<$ 8$\times$10$^{6}$
$M_{\odot}$ and Eddington ratios $\lambda_{\rm Edd}$ $\gtrsim$ 0.1, while the triple-peaked megamasers show an opposite trend.  
\end{abstract}


%%%%%%%%%%%%%%%%%%%%%%%%%%%%%%%%%%%%%%%%%%%%%%%%%%

%%%%%%%%%%%%%%%%% BODY OF PAPER %%%%%%%%%%%%%%%%%%

\section{Introduction}
Luminous 22 GHz H$_{2}$O megamaser emissions from circumnuclear environments
in active galaxies (Lo 2005) present a unique
tool to reveal the gas distribution and kinematics of active galactic nuclei (AGN) on subparsec
scales. In so-called disk maser systems, such as NGC 4258 (e.g. Herrnstein et al. 1999),
the gas resides in a subparsec scale thin disk viewed almost
edge-on (i.e. disk inclination greater than 80$^{\circ}$) and follows
nearly perfect Keplerian rotation. These disk properties
not only allow a measurement of the mass of the supermassive black
hole (BH) with an accuracy at the percent level, the geometrical/kinematic
information of a disk maser can also be modeled 
to provide a precise determination of the Hubble constant (e.g. Reid et
al. 2013; Kuo et al. 2013; Kuo et al. 2015; Gao et al. 2016).           
 
 In a typical survey of H$_{2}$O megamasers\footnote{One can find
all H$_{2}$O megamaser galaxies discovered so far and their maser
spectra on the following website : https://safe.nrao.edu/wiki/bin/view/Main/PublicWaterMaserList} 
a disk maser candidate can usually be identified if the spectrum shows the characteristic ``triple-peaked profile", i.e.
the spectrum displays three distinct maser line complexes that correspond
to the redshifted, systemic\footnote{The systemic masers refer to
the maser spectral components having velocities close to the systemic
velocity $V_{\rm sys}$ of the parent galaxy. In pristine triple-peaked maser systems, systemic masers typically have velocities within 100 km~s$^{-1}$ from $V_{\rm sys}$ (see Section 4.1).}, and blueshifted components of disk
masers (e.g. Kuo et al. 2011).         
 
 In addition to the triple-peaked sources, there are also maser galaxies
 which display only two distinct maser line
complexes. In systems such as Mrk 1210 (Zhao et al. 2018) and Circinus (Greenhill et al. 2003), the single-dish spectra reveal two distinct chunks of maser lines which are blueshifted and redshifted with respect to the recession velocities ($V_{\rm sys}$) of the galaxies, but no prominent line complexes can be seen near $V_{\rm sys}$. In these sources, there are occasionally weak maser lines arising between the redshifted and blueshifted complexes, and these lines often lie within 100 km~s$^{-1}$ from $V_{\rm sys}$ in the spectra. While one can thus define them as systemic masers, their distributions in the single-dish spectra do not allow them to be clearly distinguished from high velocity masers. It is not easy to tell whether these weak lines are simply maser features arising near the edges of the high velocity maser complexes or emission features from orbiting systemic maser clouds which reside near the line-of-sight of the central black hole as found in NGC 4258.

 Because of the lack of a maser line complex, double-peaked maser galaxies such as Mrk 1210 would not be ideal systems for determining an accurate Hubble constant at a level of $\lesssim$10\% (e.g. Reid et al. 2013; Kuo et al. 2013; Gao et
al. 2016). However, these systems could still be used to measure
BH mass ($M_{\rm BH}$) with an accuracy sufficient for constraining
the $M_{\rm BH}-\sigma_{*}$ relation (Ferrarese \& Merritt 2000;
Gebhardt et al. 2000; G$\ddot{\rm u}$tekin et al. 2009, Greene et
al. 2016, and references therein) if some of the masers arise from a rotating 
disk, such as the masers in the Circinus Galaxy (Greenhill et al. 2003).
 
 In the past decade,
the Megamaser Cosmology Project (MCP; Reid et al. 2009a; Braatz et
al. 2010) mainly focused on the triple-peaked objects to
measure $H_{0}$ and $M_{\rm BH}$ with the highest
accuracy.  As a result, the double-peaked megamasers
have been less explored.  Since there is only a small number
of case studies (Greenhill et al. 1997, 2003;
Kondratko, Greenhill, \& Moran 2005) for these systems, the physical
natures of the double-peaked megamasers are less well
understood.                                   
 
In this paper, we provide detailed imaging of
two double-peaked objects (NGC 5728 and Mrk 1) based on Very Long
Baseline Interferometry (VLBI), and we study whether or not their maser features
are associated with rotating gas disks.
Also, we provide the VLBI image and kinematics
of a triple-peaked maser galaxy, IRAS 08452$-$0011, which has a distance
of 213$\pm$15 Mpc\footnote{This is the 3K CMB distance adopted from the NASA/IPAC
Extragalactic Database (NED) in its classic form, which uses $H_{0}$ $=$ 73 km~s$^{-1}$~Mpc$^{-1}$ for evaluating the Hubble distance.}. Studying a distant maser system
like this is valuable, because such galaxies are in the Hubble
flow such that peculiar motions which
contaminate $H_{0}$ determinations become negligible. Furthermore,
such sources allow us to check whether the H$_{2}$O megamaser technique
can be applied to galaxies beyond 200 Mpc for $H_{0}$ and $M_{\rm
BH}$ determination with present sensitivity and angular resolution. Note that there are currently no direct BH mass measurements for the three maser galaxies studied here. In addition, no stellar bulge velocity dispersions ($\sigma_{*}$) are available from the literature for these sources that can enable us to infer their $M_{\rm BH}$ with the $M_{\rm BH}-\sigma_{*}$ relation. The maser distributions and kinematics presented in this paper will allow us to provide the first constraints on the BH masses for these three maser galaxies.   
  
In section 2, we present our sample of galaxies, VLBI observations
and data reduction. In Section 3, we show the VLBI images
and position-velocity diagrams of the three megamaser galaxies, 
followed by a discussion of the 
nature of two double-peaked systems.  We also
measure the BH mass for the distant triple-peaked megamaser. In
section 4, we will investigate the physical causes that lead
to different spectral characteristics.  A summary of our results
is presented in section 5.

 \section{THE SAMPLE, OBSERVATION, AND DATA REDUCTION}

\subsection{The Maser Galaxy Sample} 

\begin{figure*} 
\begin{center} 
%\vspace*{-0.3 cm} 
\hspace*{-1 cm} 
\includegraphics[angle=0, scale= 0.6]{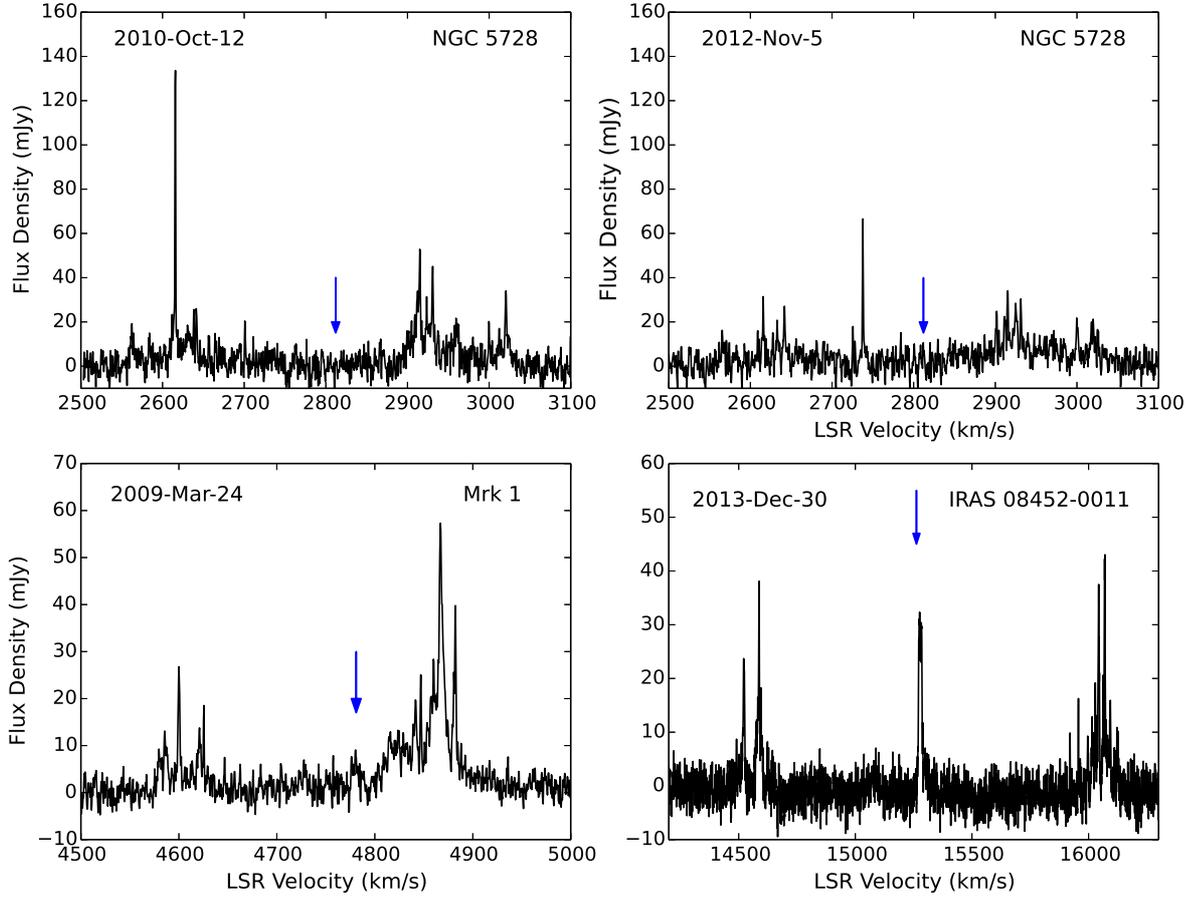} 
\vspace*{0.0 cm} 
\caption{The representative GBT spectra of the H$_{2}$O masers in
the three galaxies we studied in this paper. The date on which each
spectrum was taken is shown in the top-left corner of each panel.
The blue arrow in each panel indicates the recession velocity of the galaxy. Velocities are shown with respect to the
LSR and are based on the optical definition.  }                          
\label{figure:3maser_spectra} 
\end{center} 
\end{figure*}

The maser galaxies we study in this paper include NGC 5278, Mrk 1,
and IRAS 08452$-$0011. The H$_{2}$O masers in Mrk 1 and NGC 5728
were first discovered in maser surveys conducted by Braatz et al.
(1994) and Braatz et al. (2004), respectively, whereas water maser
emission in IRAS 08452$-$0011 was discovered in a survey as part
of the MCP in 2013. Table \ref{table:basic_galaxy_properties} lists the coordinates,
recession velocities, spectral and morphological types for these
galaxies. Their maser spectra measured with the 100-m Green
Bank Telescope (GBT)\footnote{The GBT is a facility of the Green
Bank Observatory (GBO), which is operated by the Associated Universities,
Inc. under a cooperative agreement between the National Science Foundation
(NSF) and the Associated Universities, Inc.} are shown in Figure \ref{figure:3maser_spectra}.    
 
The left two panels of Figure \ref{figure:3maser_spectra} show the representative spectra of
NGC 5728 and Mrk 1. Both spectra display two distinct line complexes
that are blueshifted and redshifted with respect to the recession
velocities of the galaxies as indicated with blue arrows.  These are
candidates for BH mass measurements, because such systems could
be triple-peaked disk maser systems with weak systemic masers
(e.g NGC 6323; see Kuo et al. 2011). 

The top-right panel 
of Figure \ref{figure:3maser_spectra} shows the maser spectrum of NGC 5728 taken 10 days before
our VLBI observation; there is a prominent 
feature at velocity 2738 km~s$^{-1}$ ($\sim$60 mJy) which could be
a systemic maser.  This line arises sporadically from time to time
over our $\sim$10 year monitoring of this source. 
The bottom-right panel shows the spectrum of IRAS 08452$-$0011, which
clearly shows the triple-peaked profile of a Keplerian
disk maser. 
This galaxy has a larger recession velocity than 
the majority of triple-peaked H$_{2}$O masers discovered
so far (see Table 1 in Kuo et al. 2018).

\begin{table*}
	\centering
        \captionsetup{justification=centering}	
	\caption{The Megamaser Sample }
	\label{table:basic_galaxy_properties} 
	\begin{tabular}{lccccrccc} % four columns, alignment for each
		\hline\hline
		 & R.A. & Decl. & $\delta$RA & $\delta$DEC & V$_{\rm sys}$ &  Distance & Spectral & Hubble    \\
	Name & (J2000) & (J2000) & (mas) & (mas) &  (km~s$^{-1}$) & (Mpc) & Type & Type \\	 
		\hline
NGC 5728   & 14:42:23.8723773$^{\rm a}$ & $-$17:15:11.015642$^{\rm a}$ & 0.06 & 0.05 & 2812 & 41.5$\pm$2.9 &Sy 1.9  & Sb\\ 
Mrk 1   & 01:16:07.2093243$^{\rm b}$ &  $+$33:05:21.633601$^{\rm b}$ & 0.12  & 0.22   & 4781    & 61.5$\pm$2.3  &Sy 2    &  S?   \\ 
IRAS 08452$-$0011   & 08:47:47.6931960$^{\rm c}$ &  $-$00:22:51.281955$^{\rm c}$ & $<$0.01 & 0.01    & 15262   & 213.3$\pm$14.9  &Sy 2    & Sa 	\\	
                \hline
\multicolumn{9}{|p{15 cm}|}{Notes. Col(1): Galaxy name; Col(2) \& Col(3) : The absolute
positions of the maser emissions determined from our phase-referencing
VLBI observations; Col(4) \& Col(5) : The position uncertainties
of the phase calibrators used to measure the absolute positions of
the maser emissions; Col(6) : The LSR velocity of the galaxy from
from NASA/IPAC Extragalactic Database (NED); Col(7): The Hubble distance
with respect to the CMB frame. Note that the values shown here are adopted from the NED in its classic form, which adopts $H_{0}$ $=$ 73 km~s$^{-1}$~Mpc$^{-1}$ for evaluating the Hubble distance; Col(8): The activity type; Col(9)
: The morphological classification provided by NED.}\\

\multicolumn{9}{|p{15 cm}|}{$\rm ^a$ The position of the maser spot at V$_{\rm op} =$
2738.0                                                                  
kms$^{-1}$, where V$_{\rm op}$ is the ``optical'' velocity of the
maser spot relative to the Local Standard of Rest (LSR).}\\  
\multicolumn{9}{l}{$\rm ^b$ The position of the maser spot at V$_{\rm op} =$
4866.0                                                                  
km~s$^{-1}$ from the phase-referencing observations BK163D (see Table
2). }\\   
\multicolumn{9}{l}{$\rm ^c$ The position of the maser spot at V$_{\rm op} =$
16043.8                                                                 
km~s$^{-1}$. }\\                 
	\end{tabular}

\end{table*}

\begin{table*}
	\centering
	\caption{Observing parameters}
	\label{table:observing_parameters} 
	\begin{tabular}{lllllcccccc} % four columns, alignment for each
		\hline\hline
Experiment   &    &  &      & Synthesized Beam    & Sensitivity      & Observing  & Phase &  $\delta\theta$ \\                                                                      
Code          & Date    & Galaxy   & Antennas$^{\rm a}$ & (mas x mas,deg)$^{\rm b}$  & (mJy)     & Mode$^{\rm c}$  & Calibrator & (Degrees)$^{\rm d}$ \\
		\hline
BB313AD    & 2012 Nov 15  & NGC 5728 & VLBA, GB  & 2.69$\times$0.38, 
$-$12.8  &  1.7  & Phase-ref.  &  J1445$-$1629  &  1.1 \\ 
BK163D   & 2010 Aug 8  & Mrk 1 & VLBA     & 0.96$\times$0.40,$-$16.9 
& 5.3  & Phase-ref.  &  J0112+3522  & 2.4  \\ 
BB261E   & 2009 Mar 29  & Mrk 1 & VLBA, GB, EB     &  1.09$\times$0.38,$-$14.5
& 1.0  & Self-cal.    &  --- & ---  \\ 
BB348B0   & 2014 Oct 5  & IRAS 08452$-$0011 & VLBA, GB      & 1.47$\times$0.69,
$-$7.5& 3.8  & Phase-ref. & J0839+0104 & 2.5   \\ 
BB348B2   & 2015 Jan 20  &  IRAS 08452$-$0011 & VLBA, GB      & 1.20$\times$0.52,
$-$4.6& 3.4  & Phase-ref. &  J0839+0104 & 2.5\\	
                \hline
\multicolumn{7}{|p{15 cm}|}{$\rm ^a$ VLBA: Very Long Baseline Array; GB: The Green Bank 
Telescope of NRAO; EB: The Effelsberg 100-m telescope  }\\  
\multicolumn{7}{|p{15 cm}|}{$\rm ^b$  Major and minor axis of the 
        synthesized beam and position angle.  }\\ 
\multicolumn{7}{|p{15 cm}|}{$\rm ^c$ ``Self-cal.'' means that the observation was conducted
in the ``self-calibration'' mode and ``Phase-ref.'' means that we
used the                                                                
``phase-referencing'' mode of observation (see section 2.2).  }\\ 
\multicolumn{7}{|p{15 cm}|}{$\rm ^d$  The angular separation between the phase calibrator
and the target source. }\\ 
        	\end{tabular}

\end{table*}

\subsection{Observations} 

The megamaser galaxies in our sample were observed between 2009 and
2015 with the Very Long Baseline Array (VLBA)\footnote{The VLBA is
a facility of the National Radio Astronomy Ob- servatory, which is
operated by the Associated Universities, Inc. under a cooperative
agreement with the National Science Foundation (NSF).}, augmented
by the GBT and in one case by the Effelsberg 100-m telescope\footnote{The
Effelsberg 100-m telescope is a facility of the Max-Planck-Institut
f$\ddot{\rm u}$r Radioastronomie}. Table \ref{table:observing_parameters} lists basic information 
regarding the observations.

We observed the megamasers either in a phase-referencing or self-calibration
mode. With phase-referencing we perform rapid switching of the telescope
pointing between the target source and a nearby ($<$ 2.5$^{\circ}$)
phase calibrator (every $\sim$50 seconds) to correct phase variations
caused by the atmosphere. The
phase calibrators used in our phase-referencing observations and
their angular separations from the target sources are listed in Table \ref{table:observing_parameters}. The absolute positions of the three
maser sources derived from our phase referencing observations are
shown in Table \ref{table:basic_galaxy_properties}.   

In a self-calibration observation, we use
the brightest maser line(s) to calibrate the atmospheric phase.
This removes the need to rapidly switch sources, improving calibration
and increasing on-source time.                                                     
For all observations, we placed ``geodetic" blocks at
the beginning and end of the observations to solve for atmosphere
and clock delay residuals for each antenna (Reid et al. 2009b). 
We also observed a nearby delay calibrator
about every 50 minutes to calibrate the single-band delay caused
by the electronic phase offsets among and across intermediate frequency
(IF) bands. In addition, a strong continuum source was 
observed in each track to calibrate the bandpass shape.               

We observed  four IF pairs for NGC 5728 in dual circular polarization with IF bands of 16 MHz and channel 
spacings of 62.5 kHz. Two IF bands were used to cover the blueshifted and  
redshifted maser line complexes, respectively, and one IF band for the $\sim$60 mJy
maser feature at 2738 km s$^{-1}$. The IF band which does not cover any maser emission was used to detect the radio continuum from the nucleus of NGC 5728. For Mrk 1, we observed four IF pairs in the phase-referencing observation and two IF pairs in the self-calibration observation. Two IF bands were used to cover the blueshifted and
redshifted masers, respectively. In the case of IRAS 08452$-$0011, we observed two IF pairs in dual
polarization with a new (Mark 5C) recording system, which allowed us to
cover the entire maser spectrum with two 128 MHz bands.  Using
the ``zoom-band'' mode of the DiFX correlator (Deller
et al. 2007), we achieved spectral channels of 24 kHz for the maser
data by re-correlating five 24 MHz sections in this manner. 
    
\subsection{Calibration}                             
We calibrated the data using the NRAO Astronomical Image Processing System (AIPS). Since the geodetic dataset and the maser dataset were taken with different frequency setups, we reduced the geodetic dataset first, then transferred solutions of atmosphere and clock delay residuals to the maser dataset.

For the geodetic dataset, we first calibrated the ionospheric delays using total electron content measurements (Walker \& Chatterjee 2000) and the Earth Orientation Parameters (EOPs) in the VLBA correlators with the EOP estimates from the US Naval Observatory (http://gemini.gsfc.nasa.gov/solve save/usno finals.erp). We performed fringe fitting with the AIPS task ``FRING" to determine the phases, single-band delays, and fringe rates of IF bands of each antenna for every geodetic source, followed by determining the multi-band delay of each antenna from these solutions. Finally, we measured the residual tropospheric delay and clock errors for all antennas using the multi-band delays. We then applied these corrections to the maser dataset.

For the maser data, after the initial data flagging, we corrected for ionospheric delay and the EOPs in the same way as for the geodetic dataset. We then corrected the sampler bias in the 2-bit correlator. The amplitude calibration was done with the information in the gain table and the system temperature table. We corrected the interferometer delays and phases caused by the effects of diurnal feed rotation (parallactic angle), and applied the tropospheric delay and clock corrections obtained from the geodetic data afterwards. The next step was to perform fringe fitting on one scan of the delay calibrators to calibrate the electronic phase offsets among and across IF bands (i.e. the single-band delay calibration). When no good single-band delay solutions can not be obtained for all IF bands of an antenna using a particular scan, we perform fringe-fitting again on another scan to obtain proper solutions for this antenna. The frequency axes of the maser interferometer spectra were then shifted to compensate for the changes in source Doppler shifts over the observing
tracks.

The final step in calibration was to solve for the atmospheric phase variation by using either phase-referencing or self-calibration. In phase-referencing mode, we ran the AIPS task CALIB on the phase calibrator to determine the phase correction as a function of time for each individual IF band. In self-calibration mode, which was adopted only for Mrk 1, we averaged multiple redshifted maser lines in the narrow velocity range of 4862$-$4871 km~s$^{-1}$ to perform phase calibration. The solution interval adopted in the self-calibration was 100 seconds. 
After the above calibrations, we discarded the phase solutions and the maser data in the time intervals within which the solutions appeared to be randomly scattered in time. The phase solutions were then interpolated and applied to all the maser data. Note that we use CALIB instead of FRING to derive the phase solutions is simply because the multiband delays in the maser dataset have been removed after we applied the tropospheric delay and clock corrections obtained from the geodetic data, and the solution interval we adopted in running CALIB is short enough to fully catch the residual atmospheric phase variation without the need of the rate information. As a result, the global fringe-fitting is not necessary for the maser dataset, and we simply derive the phase solutions with CALIB.

For IRAS 08452$-$0011, the calibrated visibility data from the two phase-referencing tracks were combined in u-v space with the AIPS task DBCON before making images for this source. Note that although the two observing tracks were conducted $\sim$3 months apart, we find no evidence that time variability of maser emission introduces noticeable errors in imaging with the combined data. Our preliminary imaging before data combination suggests that maser positions measured from the two observing tracks are consistent with each other. Given the position consistency, we further combine the visibility data for imaging to increase the sensitivity of the maser detection and enhance the accuracy of the maser position measurement.

For Mrk 1, the only target with both phase-referencing and self-calibration tracks, we only use the self-calibration track to generate the maser image presented in Section 3. Here, the role of the phase-referencing observation for Mrk 1 is to obtain the absolute position of the target source, which was then used as the phase-reference center when we calibrated the self-calibration track. The sensitivity of the phase-referencing track of Mrk 1 is four times lower than the self-calibration track, and we made $\ge$5 $\sigma$ detections of maser lines in two narrow velocity ranges of 4863$-$4867 km~s$^{-1}$ and 4914$-$4916 km~s$^{-1}$ in the phase-referencing observation. The strongest maser line was detected at 10$\sigma$ at 4866 km~s$^{-1}$ and we adopt the position of this line as the phase-reference center for Mrk 1.

After the calibration procedure described above, we Fourier transformed the gridded (u,v) data to make images of the masers in all spectral channels of the IF bands that showed maser lines, and we deconvolved the images using CLEAN with the natural weighting scheme. We fitted the detected maser spots with elliptical Gaussians to obtain the positions and flux densities of individual maser components. The measured velocity, position, and peak intensity of each individual maser spot detected in NGC 5728, Mrk 1, and IRAS 08452$-$0011 are shown in Appendix B (Tables \ref{table:NGC5728data}, \ref{table:Mrk1data}, and \ref{table:J0847data} for NGC 5728, Mrk 1, and IRAS 08452$-$0011, respectively).

\section{Results}

\subsection{VLBI Images and Position$-$Velocity Diagrams} 

\begin{figure*}
\begin{center} 
\vspace*{-0.5 cm} 
\hspace*{-1 cm} 
\includegraphics[angle=0, scale= 0.65]{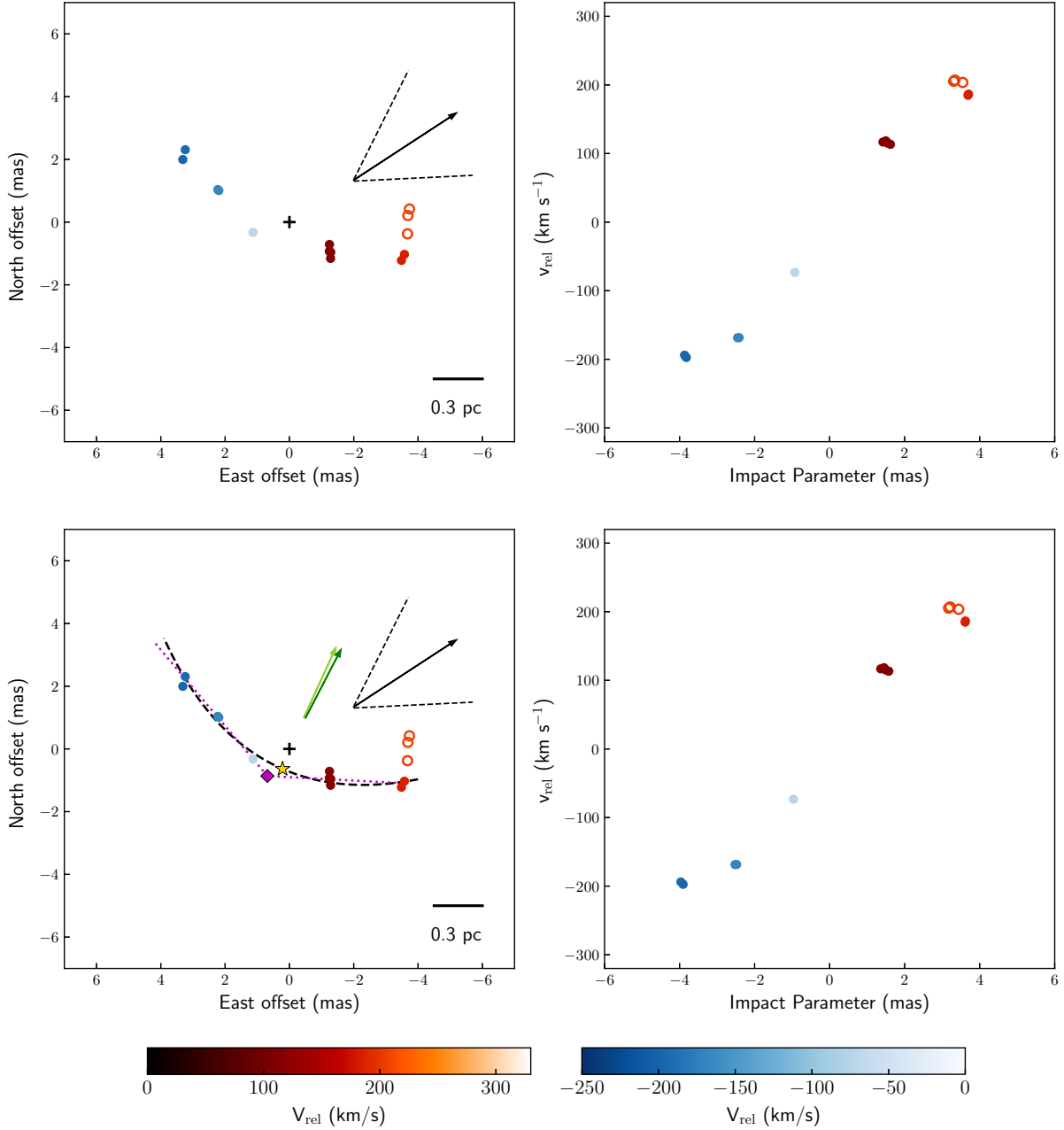} 
\vspace*{-0.9 cm} 
\caption{{\bf Top-left Panel :} The VLBI map for the 22 GHz H$_{2}$O masers
in NGC 5728. The maps are color-coded to indicate redshifted (the
red dots), blueshifted (the blue dots), and the (tentative) systemic (the
pale blue spot) masers. The open red circles show the maser spots which
are likely to be the outflow component of the maser system. The cross marks the unweighted average position of all masers spots, which is tentatively assumed to be the dynamical center for assessing the dynamical properties of the system. The black
arrow represents the projected axis of the northwestern ionization
cone reported in Wilson et al. (1993). The dashed lines indicate the opening angle ($\sim$60$^{\circ}$) of the ionization cone. The error bars on the maser
positions are not drawn here. Their values are either comparable or smaller
than the symbol size. {\bf Top-right Panel :} The position$-$velocity
diagram of the maser system. The vertical axis shows the line-of-sight maser velocity
relative to the recession velocity of the galaxy shown in Table
1. The horizontal axis shows the impact parameter relative to the assumed dynamical center. {\bf Bottom-left Panel :}  The same VLBI map for masers in NGC 5728 as in the top-left panel. The black dashed line and the magenta dotted line represent the best-fit quadratic function and cone for the maser spots shown by the solid symbols, respectively. The yellow star and magenta diamond mark the apices of the best-fit curves. The yellow-green and green arrows show the symmetric axes of the best-fit quadratic curve and cone, respectively, which indicate the outflowing direction of the wide-angle rotating wind. The opening angle of the wind is $\sim$130$^{\circ}$. {\bf Bottom-right Panel:} The position$-$velocity diagram of the maser system, with the impact parameter measured with respect to the apex of the quadratic curve.  }                                                                   
\label{figure:NGC5728_plot} 
\end{center} 
\end{figure*}

Figures \ref{figure:NGC5728_plot}, \ref{figure:Mrk1_plot}, and \ref{figure:J0847_plot} show the VLBI images and the position-velocity
(P$-$V) diagrams of the H$_{2}$O masers in NGC 5728, Mrk 1, and IRAS
08452$-$0011, respectively.  The VLBI maps
and P$-$V diagrams are color-coded to indicate redshifted (red
color), blueshifted (blue color), and systemic masers (green color).
We impose a 5$\sigma$ flux density cuffoff with respect to the flux uncertainty of each individual channel
for maser spots in NGC 5728, Mrk 1, and IRAS 08452$-$0011, respectively.
For NGC 5728 and Mrk 1, the position uncertainties of the maser spots
are either comparable to or smaller than the symbol size.             
 
Due to the lack of systemic maser components
in the double-peaked maser systems, the maser maps of NGC 5728 and
Mrk 1 are dominated by the redshifted and blueshifted masers. The
absence of the systemic masers makes it difficult to locate precisely
the dynamical centers (presumably the positions of the BHs) of these
two maser systems. For the purpose of understanding the nature and dynamical properties of these two maser systems, 
we first tentatively assume that the dynamical centers are located at the 
average position of all maser spots and for each individual maser spot we calculate the ``impact parameter", which is defined as projected radial
offset of the maser spot along the assumed disk planes relative to the dynamical centers in the VLBI maps.  
After gaining deeper insights into the nature of these maser systems, we develop methods to determine the dynamical centers more precisely by only using the high velocity masers. We present
these methods in Sections 3.3 and 3.4 for NGC 5728 and Mrk 1, respectively.

In the case of the triple-peaked source
IRAS 08452$-$0011, we locate the dynamical center by performing a
three dimensional disk modeling (e.g. Reid et al. 2013; Gao et al.
2016) of the maser system. The P$-$V diagram is then plotted relative
to the dynamical center found in the best-fit disk model.

\begin{table*}
	\centering
	\caption{Upper limits to the Continuum Emission from our Three Megamaser Galaxies }
	\label{table:continuum} 
	\begin{tabular}{lcccl} % four columns, alignment for each
		\hline\hline
Galaxy      & $\nu$$_{\rm Center}$ & $\Delta\nu$  & 
I$_{1\sigma}$  & Project Code \\ 
 & (GHz)  & (MHz) & (mJy~beam$^{-1}$)    &   \\
	\hline
NGC 5728      &   21.986565   &  12.5  &    $<$ 0.19   &  BB313AD
\\                                                                      
Mrk 1              &    21.902685    &   5.0  &   $<$ 0.13    & BB261E
\\                                                                      
IRAS 08452$-$0011 &     21.193950    &  22.5   &   $<$ 0.14    &
BB348B0, BB348B2      \\
              \hline     
 \multicolumn{5}{|p{10 cm}|}{Note. Col(1): Name of the galaxy; Col(2): The central frequency
of spectral range covered by                                            
line-free channels used to search for continuum emission; Col(3):
The total spectral width of the selected maser line-free channels;
Col(4): The $1\sigma$ detection limit of the continuum                  
emission; Col(5): The data used for continuum detection. }\\              
	\end{tabular}

\end{table*}

\subsection{Search For Continuum Emission}
 
We searched for continuum emission from the vicinities of the supermassive
BHs in our megamaser galaxies by
averaging the line-free spectral channels in our data and imaging
with natural weighting to maximize the detection sensitivity. We
detected no continuum emission in all megamaser galaxies presented
in this paper. The central frequencies and the total spectral widths
of the selected maser line-free channels used for channel averaging
as well as the continuum upper limits are listed in Table \ref{table:continuum}.

\subsection{The Nature of the Maser System in NGC 5728} 
\subsubsection{Are the masers in a wind ?}
NGC 5728 is an active galaxy that hosts a Seyfert 2 nucleus. It is known
for its spectacular biconical ionization cone seen in the optical
emission line images (Wilson et al. 1993), which has an extent of 1.8 kpc. 
Such an ionization cone (Durr$\acute{\rm e}$ \& Mould 2018, 2019) is thought
to be caused by AGN winds driven by radiation pressure or winds from
the accretion disk, with the axis of the bi-conical cone structure
co-aligned with the rotation axis of the accretion disk. The evidence
for the ionization cone tracing a bipolar outflow can be seen in
the kinematics of the gas in the bicone (Durr$\acute{\rm e}$ \& Mould
2019), which show clear signatures of acceleration in the velocity
maps and position-velocity diagrams.
 
Wilson et al. (1993) reported that the
apex of the ionization cone, which reflects the location of the obscured
active nucleus, is at the same location as the radio nucleus\footnote{The
coordinate of the nucleus position reported in Schommer et al. (1988)
is $\alpha$(J2000) $=$ 14:42:23.884$\pm$0.015, $\delta$(J2000) $=$
$-$17:15:10.81$\pm$0.20} measured by Schommer et al. (1988). This
position agrees with that of the masers (see Table \ref{table:basic_galaxy_properties}) measured by our 
VLBI observation, consistent with the picture that the H$_{2}$O
maser disk is at the center of the AGN and is surrounded by
an obscuring torus (Masini et al. 2016).                                
 
The top-left panel of Figure \ref{figure:NGC5728_plot} shows the H$_{2}$O maser distribution
in NGC 5728. The black arrow in the plot represents the projected
axis of the northwestern ionization cone reported in Wilson et al.
(1993) whereas the two dashed lines indicate the opening angle ($\sim$60$^{\circ}$; Wilson et al. 1993) of the ionization cone. 
As one can see in this plot, except for the three maser spots
shown by open red circles, the maser distribution follows a curved
distribution which resembles a nearly edge-on warped maser disk.
The three open circles may represent outflowing maser spots lifting
off from the disk. If our disk interpretation is correct, the inner
and outer radii of the disk are 0.38 pc and 0.82 pc, respectively,
and the position angle of the  disk (P.A.)\footnote{The position
angle (P.A.) increasing counterclockwise is calculated from the slope
of the straight line fit to the blueshifted and redshifted components
of the disk. P.A. $=$ 0$^{\circ}$ when the redshifted side of the
disk is oriented northwards.} is 244 degrees. The size of the
disk would be consistent with the typical size of maser disks
shown in Gao et al. (2017).                                             
 
While the maser distribution in NGC 5728 is consistent with masers
being in a disk, the ``rotation curve" is not
similar to that of other maser disks
(e.g. Greenhill et al. 1997, 2003, Reid et al. 2009a; Kuo et al.
2011; Gao et al. 2017, Zhao et al. 2018). In the top-right panel of Figure
2, we show the maser velocity as a function of the impact parameter,
which is defined as projected (cylindrical) distance of a maser spot
relative to the rotation axis. Here, the rotation axis is defined
as the axis perpendicular to the assumed disk plane indicated by
the line best-fit to the positions of the blueshifted and redshifted
masers without including the maser spots represented by the open
circles.                                                                
From this plot, it can be seen that instead of falling with 
increasing magnitude of the impact parameter
as seen for Keplerian rotation, the velocities of the blueshifted and redshifted
masers increase with their projected radii.  If the gravity of the central supermassive
BH dominates the dynamics of the maser spots and the maser spots
reside in a thin disk, then the rising rotation curve could 
be explained by line-of-sight projections of maser spots confined to
a narrow annulus, just like the systemic masers in NGC 4258 (Herrnstein et
al. 1999). However, long velocity-coherent path lengths, necessary
for strong maser amplification are not favored for such a configuration.
                                                       
We note that a thin annulus in a Keplerian system is not
the only way to produce a rotation curve that rises linearly with
radius.  An alternative model that can give rise to the solid-body like
rotation seen in NGC 5728 is that the masers could originate in a wide-angle
magnetocentrifual wind (e.g. Blandford \& Payne 1982; Proga 2000;
Ouyed \& Pudritz 1997; Ustyugova et al. 1999; Krasnopolsky, Li, \&
Blandford 1999) which is launched from a disk. 
In the example of the young stellar object (YSO) Orion Source-I (e.g. Matthews et al. 2010; Goddi et al. 2011;
Greenhill et al. 2013), which shows evidence for the presence
of a magnetocentrifugal disk wind, the outflowing gas traced by Si$^{18}$O
line emission also shows a linearly rising velocity as a function
of impact parameter (see Figure 2 in Hirota et al. 2017). This suggests
that the wind is corotating with the gas ring in the circumstellar
disk (Hirota et al. 2014).                                              
 
Wind rotation is expected in a magnetocentrifugal disk wind (e.g.
Proga 2000; Ouyed \& Pudritz 1997; Ustyugova et al. 1999; Krasnopolsky,
Li, \& Blandford 1999) because the magnetic field lines threading
the outflowing gas are anchored to a rotating disk. This leads to a
well-known feature of this type of wind $-$ the conservation
of the specific angular velocity of the gas (Proga 2000). That is,
if a maser clump is outflowing from a foot point in the mid-plane
of a Keplerian disk at radius is $r_{0}$ and Keplerian
velocity is $v_{\rm k}$, the rotational velocity will be $v_{\phi}$
$=$ ($r'$/$r_{0}$)$v_{\rm k}$ when it reaches a
radius of $r'$  (Kashi et al. 2013).  
In such a situation, a linearly rising rotation curve can arise
when the outflowing gas follows roughly the same magnetic
streamlines. Alternatively, such a rotation curve can also appear 
if the outflow is launched from a thin annulus in a Keplerian disk as in the case of the Si$^{18}$O outflow in Orion Source I (Hirota et al. 2017).
  
%The overall
%velocity extent of the NGC 5728 masers is small ($\lesssim$400
%km~s$^{-1}$) compared to most Keplerian maser disk systems ($\sim$1000
%km~s$^{-1}$; Kuo et al. 2011; Gao et al. 2016, 2017; Zhao et al.
%2018), suggesting that the masers are located further away from the
%nucleus than we see for accretion-disk masers.  Also, 
  
In addition to providing an explanation for a rising rotation curve,
a magnetocentrifugal wind can explain other data. 
The maser distribution on the sky displays a curved shape with both ends 
pointing toward the same direction.  This is inconsistent with
a symmetrical warping as is common in disk masers.
The NGC 5728 configuration is reminiscent of the masers in Circinus
(Greenhill et al. 2003), where only one side of the outflow is 
observed and the other side is thought to be blocked by the warped disk.
Moreover, the masers in NGC 5728 roughly form a cone as in Circinus (see the bottom-left panel of Figure \ref{figure:NGC5728_plot}), with the symmetric axis of the cone (see the green arrow) aligning with the axis of the northwestern ionization cone within $\sim$30$^{\circ}$. It is likely that the wide-angle (opening angle $\sim$130$^{\circ}$) wind traced by maser emission at $\sim$1 pc scale resides at the base of the kpc-scale ionization cone (opening angle $\sim$60$^{\circ}$), with the wind flowing in a similar direction as the gas in the ionization cone (see the green arrow in the plot). 

We note that such a magnetically driven wind with a wide opening angle of $\sim$130 $^{\circ}$ is theoretically possible in the context of the disk wind models proposed by Blandford \& Payne (1982) and K\"{o}nigl \& Kartje (1994). The half-opening angle of the streamlines along which the masers follow (i.e. $\sim$65$^{\circ}$) satisfies the necessary condition ($\theta_{p}$ $>$ 30$^{\circ}$)\footnote{$\theta_{p}$ indicates the angle between the poloidal component of the magnetic field along which gas flows and the polar axis of the wind. } for magnetocentrifugal wind generation as found by Blandford \& Payne (1982). The A2 model studied in K${\rm \ddot{o}}$nigl \& Kartje (1994) allows a centrifugally driven hydromagnetic wind to have a full opening angle of 124$^{\circ}$ if the AGN bolometric luminosity is $L_{\rm bol}$$\sim$10$^{45}$ erg~s$^{-1}$. For NGC 5728, $L_{\rm bol}$ inferred from [OIII] luminosity is (6.0$^{+8.5}_{-3.5}$)$\times$10$^{44}$ erg~s$^{-1}$, suggesting that this maser galaxy would have sufficiently high AGN luminosity to engender a wide-angle wind. As indicated by Blandford \& Payne (1982), a magnetically driven wind will get collimated by the toroidal component of the magnetic field as the wind reaches larger distances from the disk. As a result, one could speculate that the opening angle of the maser wind in NGC 5728 may reduce substantially at larger scales and the wind may gradually merge with the gas in the kpc-scale ionization cone.

%For NGC 5728, this would imply that the northwestern ionization
%cone is coming towards us. Indeed, when one examines the gas kinematics
%in the northwestern ionization cone, one can see that the gas is
%blueshifted (see Figure 3 \& 4 in Durr$\acute{\rm e}$ \& Mould
%2019).

%, supporting the picture that the masers in NGC 5728 are part
%of a bipolar outflow approaching toward the observer.                 

\subsubsection{Estimating the black hole mass} 
To estimate the BH mass ($M_{\rm BH}$) of this maser system that has non-Keplerian
kinematics, we note that the rotational velocities of the masers will be $v_{\phi}$
$=$ ($r'$/$r_{0}$)$v_{\rm k}$ if the masers follow a magnetocentrifugal wind and corotate with the disk as we suggested for NGC 5728. Before using the rotational velocities of the masers to evaluate the BH mass, a more precise determination of the dynamical center is important for minimizing the systematic uncertainty of the BH mass estimate. To do so, we note that in Circinus, the masers in the wind, although asymmetric, roughly form a cone with the apex consistent with the dynamical center. In addition, for the well-known YSO Orion Source I, the boundaries of the rotating winds represented by the SiO (Matthews et al. 2010) and Si$^{18}$O emissions (Hirota et al. 2017) also form a smooth, roughly axisymmetric curve with the apex coinciding with the dynamical center. This suggests that fitting a cone or a smooth/axisymmetric curve (e.g. a quadratic function) to the high velocity masers in NGC 5728 could allow one to better locate the dynamical center in this maser system. 

In the bottem-left panel of Figure \ref{figure:NGC5728_plot}, we show the quadratic function (the thick dashed line) and cone (the dotted line) that are best fit to the maser spots represented by the solid symbols. The apices of the curve and cone are shown by the yellow star and magenta diamond, respectively. Adopting the position of the yellow star as the dynamical center, we measure the impact parameters relative to the new center along a plane perpendicular to the symmetric axis of the best-fit curve.  The new P$-$V diagram is plotted in the bottom-right panel of Figure 2. Note that the P$-$V diagram would only have a small change (i.e. the impact parameter increases by 0.34 mas for both blue and red dots) if we adopt the apex of the cone as the dynamical center.

Measuring the BH mass for the wind maser system in NGC 5728 requires knowledge of $r_{0}$ and $v_{\rm k}$, which represent the radius and orbital velocity of the launching point of an outflowing maser clump in the disk.  We note that the innermost blueshifted maser spot is likely to lie close to the disk plane, and its enclosed mass ($M_{\rm enc}$ $=$ 2.4$\times$10$^{5}$ $M_{\odot}$) could be used to estimate $M_{\rm BH}$. However, we also note that the impact parameter ($r$ $=$ 1.0 mas) of this tentative systemic maser feature is smaller than the dust sublimation radius ($r_{\rm sub}$ $=$ 1.6 mas or 0.3 pc)\footnote{The sublimation radius is estimated with Equation (1) in Netzer (2015) for silicate grains and the bolometric AGN luminosity of NGC 5728 shown in Table \ref{table:luminosities}.  }. Given the expectation that H$_{2}$O masers cannot occur within $r_{\rm sub}$, this suggests that this maser spot may not reside close to the mid-line of the disk, and we thus avoid using this maser spot to estimate $M_{\rm BH}$. For other maser spots in the system, the values of  $r_{0}$ and $v_{\rm k}$ cannot be inferred precisely based on the maser map and the P$-$V diagram since the streamlines along which the outflowing masers follow cannot be well-constrained.

Nevertheless, one can still place a conservative upper bound for the BH mass by using the outermost (blueshifted) maser spot. Since the gas is flowing out in a wind that conserves angular velocity, one can expect that $r'$ $>$ $r_{0}$ and $v_{\phi}$ $>$ $v_{\rm k}$, where $r'$ and $v_{\phi}$ represent the radius and the rotational velocity of the outflowing maser spot at the observed position. This suggests that $M_{\rm BH}$ $\equiv$ $r_{0}v_{\rm k}^{2}$/G $<$ $r'v_{\phi}^{2}$/G. If the orbital plane of the gas is not exactly edge-on, the above equation needs to be revised as $M_{\rm BH}$ $\equiv$ $r_{0}v_{\rm k}^{2}$/G $<$ (sin $i$)$^{-2}$($r'v_{\phi, los}^{2}$/G), where $i$ and $v_{\phi, los}$ represent the inclination angle of the rotating plane and the orbital velocity seen along the line of sight, respectively. Adopting the dynamical center marked by the yellow star and assuming that rotating plane of the masing gas is within 20$^{\circ}$ from being edge-on (i.e. 70$^{\circ}$ $<$ $i$ $<$ 110$^{\circ}$)\footnote{All pristine Keplerian maser disks have inclination angles 80$^{\circ}$ $<$ $i$ $<$ 100$^{\circ}$ (e.g. Herrnstein et al. 1999; Kuo et al. 2011; Gao et al. 2017; Zhao et al. 2018), and this is consistent with the picture that a maser disk prefers to be nearly edge-on (i.e. $i$ $=$ 90$^{\circ}$) because such an orientation provides the longest path lengths for maser amplification. Given this fact, it would be reasonable to assume that the angular offset ($\Delta\theta_{i}$) of the orbital plane from the edge-on orientation is within 10$^{\circ}$ for NGC 5728 as well. Nevertheless, for the purpose of placing a more conservative upper limit for the BH mass in NGC 5728, we choose $\Delta\theta_{i}$ $\lesssim$20$^{\circ}$ as the upper bound for the inclination offset. This limit is five times greater than the mean angular offset ($\overline{\Delta\theta_{i}}$ $=$ 4.0$^{\circ}$) of the 12 well-modeled maser disks studied in the above-mentioned references.}, one can place an upper limit of the BH mass to be $M_{\rm BH}$ $<$ 8.2$\times$10$^{6}$ $M_{\odot}$. If we adopt the apex of the cone as the dynamical center, the upper bound of the BH mass becomes $M_{\rm BH}$ $<$ 7.5$\times$10$^{6}$ $M_{\odot}$. Here, we adopt the greater of these two values as the upper limit of the BH mass and conclude that $M_{\rm BH}$ $<$ 8.2$\times$10$^{6}$ $M_{\odot}$ for NGC 5728.

%and $M_{\rm enc}$ $\equiv$ $r'v_{\phi, los}^{2}$/G is simply the enclosed mass of an outflowing maser clump. 

\begin{figure*}
\begin{center} 
%\vspace*{-0.3 cm} 
\hspace*{-2 cm} 
\includegraphics[angle=0, scale= 0.47]{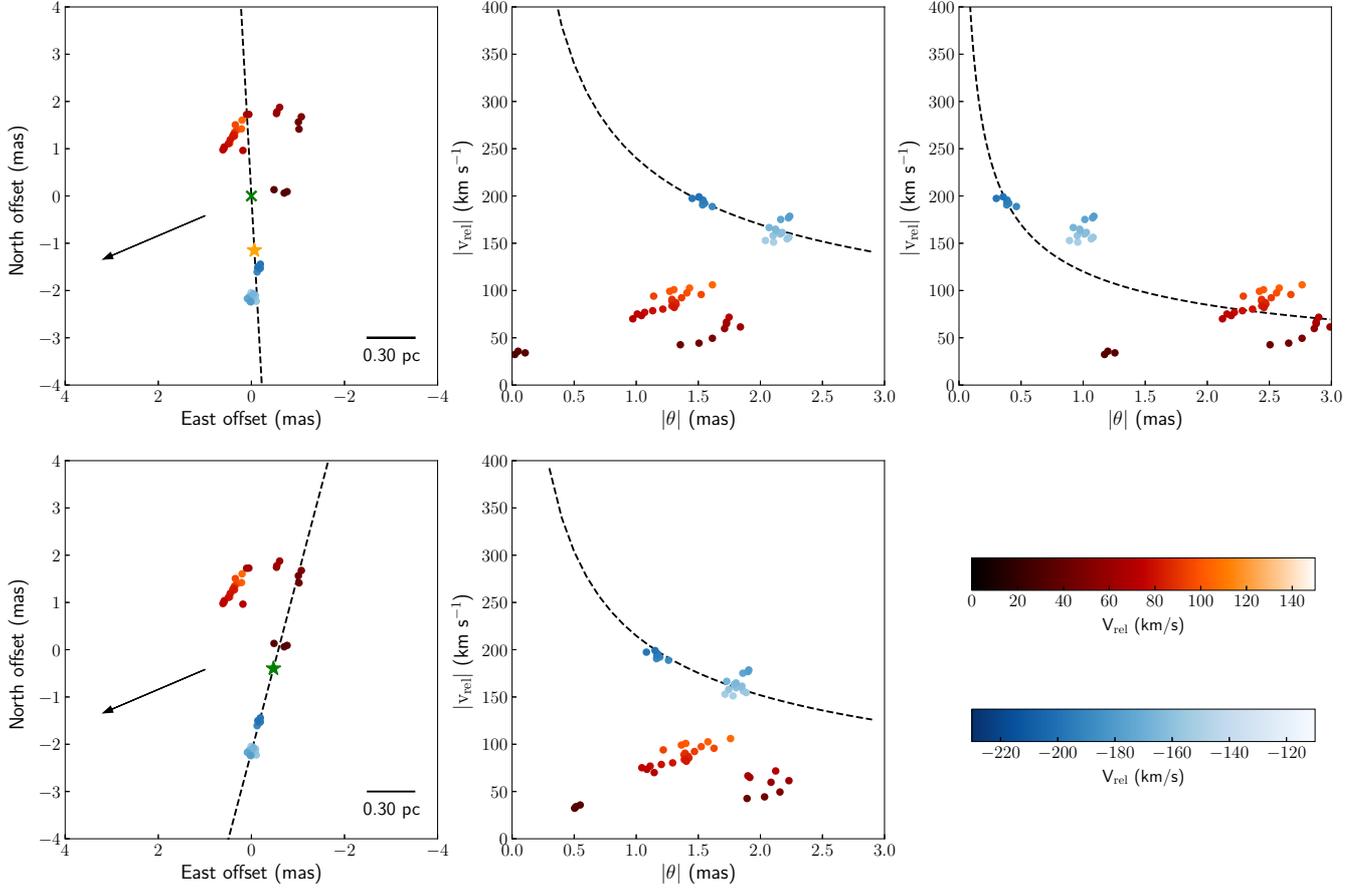} 
\vspace*{-1.0 cm} 
\caption{{\bf Top-left Panel :} The VLBI map for the 22 GHz H$_{2}$O
masers in Mrk 1. The maps are color coded to indicate red- and blueshifted
masers. The green cross marks the unweighted average position of all masers spots, which is tentatively assumed to be the dynamical center for evaluating the dynamical properties of the system.The
black arrow indicates the axis of the radio jet reported in Kamali
et al. (2019). The dashed line shows the assumed maser disk plane,
which is defined by the line joining the average positions of the
blueshifted and redshifted masers, respectively. The error bars on
the maser positions are not drawn here because their values are either
comparable or smaller than the symbol size.  {\bf Top-center Panel
:} The position$-$velocity diagram of the maser system, with the
impact parameter evaluated with respect to the assumed dynamical
center shown by the cross in the left panel. $| \theta |$ in the
horizontal axis indicates the absolute value of the impact parameter. The vertical axis shows the absolute value of the line-of-sight maser velocity ($|{\rm v_{rel} }|$) relative to the recession velocity of Mrk 1 (Table \ref{table:basic_galaxy_properties}). The dashed line shows the Keplerian rotation curve calculated with
a central BH mass of $M_{\rm BH}$ $=$ 4.0$\times$10$^{6}$ $M_{\odot}$.
{\bf Top-right Panel :}   The position$-$velocity diagram of the
maser system relative to a different dynamical center marked by the
yellow star in the left panel. The dashed line shows the Keplerian
rotation curve calculated with a BH mass of $M_{\rm BH}$ $=$ 1.0$\times$10$^{6}$
$M_{\odot}$. {\bf Bottom-left Panel :} The same VLBI map as shown in
the top-left panel. The green star indicates the BH location
inferred from fitting a Keplerian rotation curve to the blueshifted masers. The dashed line represents the newly adopted
disk plane which passes through the average positions
of the two blueshifted maser clumps.  {\bf Bottom-center Panel :} The position$-$velocity
diagram constructed relative to the assumed dynamical center shown
by the green star in the bottom-left panel. The dashed line shows
the best-fit Keplerian rotation curve to the blueshifted masers, which gives a BH mass of $M_{\rm
BH}$ $=$ (3.2$\pm$0.5)$\times$10$^{6}$ $M_{\odot}$.   }                           
\label{figure:Mrk1_plot}  
\end{center} 
\end{figure*}

\subsection{ The Nature of the Maser System in Mrk 1 : Outflow or
Perturbed Disk ?  }
                                                     
Mrk 1 is a Seyfert 2 galaxy at a distance of $\sim$62 Mpc. Analysis of
an X-ray spectrum from XMM-Newton observations (Guainazzi, Matt,
\& Perola 2005) suggests that Mrk 1 hosts a Compton-thick nucleus
where the majority of disk maser systems tend to reside (Greenhill
et al. 2008; Zhang et al. 2010; Masini et al. 2016). However, our
VLBI observation shows that the maser distribution in Mrk 1 is clearly
inconsistent with masers residing in a thin disk viewed nearly edge-on.
                                                    
As one can see in the top-left panel of Figure \ref{figure:Mrk1_plot}, while the blueshifted
masers cluster at two spatially displaced clumps, the majority of
the redshifted maser spots appear to be distributed along a curve
reminiscent of a bow shock. This could give a hint that the masers
may be part of a bipolar molecular outflow propagating along the
north-south direction, with the redshifted masers tracing the front
of the bow shock that occurs when the outflow is impinging onto
the ambient gas. However, we find that this interpretation is inconsistent
with the finding of Kamali et al. (2019), who discovered a radio
jet propagating in a direction with P.A. of $113^{\circ}\pm5^{\circ}$
(indicated by the black arrow in the left panel of Figure \ref{figure:Mrk1_plot})
over a length of $\sim$30 pc. The jet orientation is apparently far 
from being co-aligned
with the direction of the suggested gas outflow\footnote{The relative
position between the masers and the jet can be seen in Figure 1 in
Kamali et al. (2019)}.  This argues against the interpretation of an outflow
propagating in the north-south direction. 

Moreover, based on the line-of-sight velocities of the redshifted masers, the speed of the putative outflow is expected to be $\gtrsim$70 km~s$^{-1}$. Assuming a sound speed of 2.3 km~s$^{-1}$, corresponding to a temperature of 800 K (Herrnstein et al. 2005), the Mach number of the masing medium would be greater than $\sim$30. This suggests that the bow shock front would have a U shape (Furuya et al. 2000), inconsistent with the relatively blunt parabolic structure shown by the redshifted masers. Therefore, one can also rule out the scenario that the redshifted masers trace a bow shock front caused by an impinging outflow.                  
 
Instead of interpreting the masers as being part of an outflow, one
could interpret the maser distribution as a perturbed disk. In the
top-left panel of Figure \ref{figure:Mrk1_plot}, we draw a dashed line which goes through
the (unweighted) average positions of the redshifted and blueshifted
masers, respectively. If we define this line at P.A. = 3$^{\circ}$
as the plane of a thick disk (the eight redshifted maser spots
with negative eastern offsets could be outflowing gas components lifting
from the disk), then we will have a rotation axis of the disk which
is offset from the jet propagation direction by only $20^{\circ}$,
well within the offset distributions reported by Greene et al. (2013)
and Kamali et al. (2019).        
Adopting this interpretation, we assume that the dynamical
center of the disk is located at the unweighted average position of all maser
spots (green cross in the top-left panel of Figure \ref{figure:Mrk1_plot}). 
The corresponding P$-$V diagram is shown in the top-center panel of Figure \ref{figure:Mrk1_plot}.
Here, we adopt the recession velocity ($V_{\rm sys}$ $=$ 4781 km~s$^{-1}$) of Mrk 1 from NED (see Table \ref{table:basic_galaxy_properties}) and fold the velocities as well as the impact parameters of the maser spots by plotting their magnitudes.
We do so in order to better reveal the differences of the velocity distributions between the redshifted and blueshifted masers. 
 
The P$-$V diagram shows that the velocities of the
blueshifted maser spots are consistent with Keplerian rotation (the dashed line in the diagram) for a central BH mass of 
$M_{\rm BH}$ $=$ 4$\times$10$^{6}$ $M_{\odot}$.                                                            
On the other hand, the P$-$V diagram also reveals that the kinematics
of the redshifted masers are far more complicated than those of the
blueshifted masers and inconsistent with Keplerian rotation. 
Except for the three maser spots near the dynamical center, the velocity distribution of these maser spots
can be separated into two groups, each with speeds increasing 
with impact parameter. In addition, the speeds of the redshifted
maser spots are systematically lower than those of the blueshifted
masers. We find no simple model that can explain this kind of kinematic distribution.     
  
Nevertheless, we note that the peculiar velocity offsets between
the blueshifted and the redshifted masers can be reduced significantly
if one shifts the dynamical center toward the blueshifted masers. We try a range of shifts between 0 and 1.4 mas along the assumed disk plane, and find it unlikely to locate a dynamical center that allows both blueshifted and redshifted masers to be consistent with a Keplerian rotation curve. One can at best find an offset (e.g. 1.15 mas; see the yellow star in the top-left panel) that allows a Keplerian rotation curve\footnote{This rotation curve is calculated with with a BH mass of $M_{\rm BH}$ $=$ 1.0$\times$10$^{6}$ $M_{\odot}$} to fit the redshifted masers and one of the two blueshifted maser clumps (see the top-right
panel). This suggests that the redshifted masers may not really reside in a perturbed disk as we suggested, and a different scenario needs to be considered. 

Since the morphology/kinematics of the redshifted masers are highly disturbed while the structure/kinematics of the blueshifted masers are well-ordered, it is plausible to speculate that the blueshifted masers are in a disk and the majority of the redshifted masers (e.g. the redshifted masers residing on the left of the dashed line) are in a wind. Given this interpretation for Mrk 1, we re-define the disk plane as the line passing through the unweighted average positions of the two blueshifted maser clumps (the dashed line in the bottom-left panel). Since this plane goes through some redshifted maser spots, it is likely that these masers also lie close the disk plane. If true, their velocities, which are substantially closer the systemic velocity of the galaxy than the blueshifted masers, would suggest that either their positions are offset from the mid-line of the disk, or these masers reside on the equatorial plane of a larger scale torus (i.e. they lie further away from the BH) that surrounds the maser disk (e.g. Sawada-Satoh et al. 2000; Kondratko et al. 2005). 

To estimate the BH mass of this maser system, we fit a Keplerian rotation curve to the blueshifted masers and allow the position of the dynamical center to be a free fitting parameter. When fitting the rotation curve, we add an error floor of 4.2 km~s$^{-1}$ to the velocity data to account for the intrinsic velocity scatter and allow the reduced $\chi^{2}$ of the fit to be 1.0. The best-fit dynamical center, which has an uncertainty of 0.2 mas along the assumed disk plane, is marked by the green star in the bottom-left panel of Figure \ref{figure:Mrk1_plot}. Our fit gives a BH mass of (3.2$\pm$0.5)$\times$10$^{6}$ $M_{\odot}$ for the Mrk 1 maser system. Here, the BH mass error reflects the formal uncertainty of the least $\chi^{2}$ fitting. This error would only change slightly from 0.5$\times$10$^{6}$ $M_{\odot}$ to 0.51$\times$10$^{6}$ $M_{\odot}$ if we also consider the effect of disk inclination angle ($i$) in the BH mass estimation and assume that 70$^{\circ}$ $<$ $i$ $<$ 110$^{\circ}$ (see Section 3.3.2).

Note that in the rotation curve fitting described above, we assume that the effect of non-gravitational motion is negligible and the maser velocities 
are assumed to be entirely gravitationally driven. This assumption will be justified in Section 4 and Appendix A, which provide quantitative estimates of the systematic uncertainty of the BH mass caused by ignoring the effect of wind disturbances for the double-peaked maser systems discussed in this paper.

\begin{figure*}
\begin{center} 
\hspace*{-0.5 cm} 
\includegraphics[angle=0, scale= 0.68]{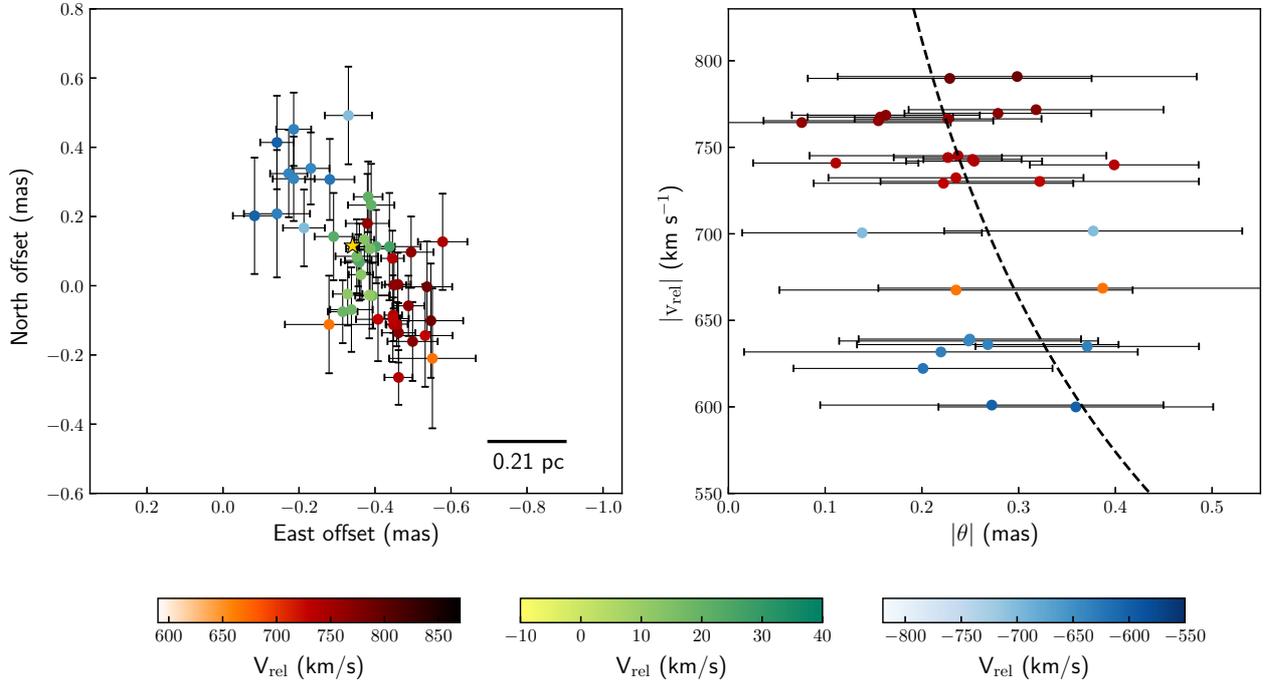} 
\caption{{\bf Left Panel :} The VLBI map for the 22 GHz H$_{2}$O
masers in IRAS 08452$-$0011. The maps are color coded to indicate
redshifted, blueshifted, and systemic masers. The yellow star in
the plot shows the best-fit position of the dynamical center obtained
from our disk modeling. {\bf Right Panel :} The rotation curve of
the disk maser system. $| \theta |$ in the horizontal axis indicates
the absolute value of the impact parameter. $|{\rm v_{rel} }|$ in the vertical axis shows the absolute value of the line-of-sight maser velocity 
relative to the recession velocity of the galaxy.The dashed line shows
the Keplerian rotation curve calculated based on the best-fit BH
mass $M_{\rm BH}$ = (3.3$\pm$0.2)$\times$10$^{7}$ $M_{\odot}$ from
the disk modeling.}                                                     
\label{figure:J0847_plot}  
\end{center} 
\end{figure*} 

\subsection{BH Mass Measurement with IRAS 08452$-$0011}
 
IRAS 08452$-$0011 is a Seyfert 2 galaxy at a distance of 213$\pm$15 Mpc (the classic NED; see Table \ref{table:basic_galaxy_properties}). 
It is one of the
three farthest disk maser candidates\footnote{The other two disk
maser candidates with $D$ $\gtrsim$200 Mpc are Mrk 34 and 2MASX J01094510$-$0332329
(or J0109$-$0332 in the MCP maser catalog) have LSR recession
velocities of 15145$\pm$90 km~s$^{-1}$ and 16363$\pm$30 km~s$^{-1}$,
respectively (the values are adopted from NED).} which lie beyond
$\sim$200 Mpc. Before our present work, the most distant disk maser
that has been imaged with VLBI is NGC 6264 (Kuo et al. 2011). The
modeling of the maser disk in that galaxy gives a distance of 144$\pm$19
Mpc (Kuo et al. 2013).                                                  
 
In the left panel of Figure \ref{figure:J0847_plot}, we show the spatial distribution of
the maser spots in IRAS 08452$-$0011, which is consistent with the
general expectation for a maser disk (e.g. Kuo et al. 2011; Gao et
al. 2017). Rather than being intrinsically thick, the apparent thickness
of the disk is caused by the relatively large position                  
uncertainties of the maser spots, which appear to be significant
in IRAS 08452$-$0011 simply because the maser disk at the distance
of 213 Mpc only has a radius of 0.5 mas and is smaller in angular size than other nearby
maser disks that have been imaged (e.g. Kuo et al. 2011; Gao
et al. 2017; Zhao et al. 2018). The position uncertainties shown
in Figure \ref{figure:J0847_plot} are actually comparable with the position errors of the
masers in NGC 5728, which are negligible compared to the size of
the maser disk shown in Figure \ref{figure:NGC5728_plot}.                                       
 
To obtain the BH mass and disk properties of this system, we adopt
a Bayesian approach to model the maser disk in three dimensions (see
Reid et al. 2013; Humphreys et al. 2013 ; Gao et al. 2016) with a
fitting code described in Reid et al. (2013). In this approach, the
disk modeling code adjusts model parameters and minimizes
the residuals of the position ($x, y$), velocity, and acceleration
for each maser spot. Global model parameters involved in the modeling
include the Hubble constant ($H_{0}$), black hole mass ($M_{\rm BH}$),
recession velocity of the galaxy ($V_{\rm sys}$), and  other parameters
that describe the orientation and warping of the disk. We summarize these parameters and the priors adopted in the disk modeling in Table \ref{table:modelparameters}.

\begin{table}
	\centering
	\caption{IRAS 08452$-$0011 H$_{2}$O Maser Model }
	\label{table:modelparameters} 
	\begin{tabular}{ lccc} % four columns, alignment for each
		\hline\hline
Parameter      & Priors & Posterioris   & Units  \\
	\hline
$H_{0}$ & 73  & ---      & km~s$^{-1}$~Mpc$^{-1}$    \\
$V_{sys}$ & --- & 15282.1$\pm$4.1 & km~s$^{-1}$               \\
$V_{cor}$ & 311    & ---      & km~s$^{-1}$               \\
$M$     &  ---          & 3.3$\pm$0.2      & 10$^{7}$~$M_{\odot}$      \\
x$_{0}$ &  ---         & $-$0.341$\pm$0.008    & mas                       \\
y$_{0}$ &  ---         & 0.115$\pm$0.019    & mas                       \\
$i_{0}$ &  ---          & 85.2$\pm$0.3         & deg                       \\
$d i/d r$ &  0.0           & ---       & deg~mas$^{-1}$             \\
$p_{0}$ & ---      & 207.3$\pm$2.6       & deg                        \\
$dp/dr$ & 0.0  & ---       & deg~mas$^{-1}$ \\
$d^{2}p/dr^{2}$ & 0.0  & ---       & deg~mas$^{-2}$ \\
              \hline     
 \multicolumn{4}{|p{8 cm}|}{Note. Parameters are as follows: Hubble constant ($H_{0}$), optically defined $V_{LSR}$
of the central black hole ($V_{sys}$), peculiar velocity with respect to Hubble 
flow in cosmic microwave background frame ($V_{cor}$), black hole mass ($M$), eastward
(x$_{0}$) and northward (y$_{0}$) position of the black hole with respect to the phase-reference center (Table \ref{table:basic_galaxy_properties}), disk inclination ($i_{0}$) and
inclination warping (change of inclination with radius: $i_{1}\equiv di/dr$), disk position
angle ($p_{0}$) and position angle warping (change of position position
angle with radius: $dp/dr$ and $d^{2}p/dr^{2}$). Flat priors were used, except where
listed. Parameter values given in this table were produced from binned histograms for each parameter and finding the center
of the central 68\% of the probability distribution. We assign the difference between the center and edge
of the central 68\% distribution to be the parameter uncertainty.  }\\              
	\end{tabular}

\end{table}

When modeling the disk, we assume that the high-velocity maser components
reside at the mid-line of the disk and have zero centripetal accelerations
along the line of sight.  In addition, we adopt a Hubble constant of $H_{0}$ $=$ 73 km~s$^{-1}$ so that the corresponding galaxy distance is consistent with the value shown in Table \ref{table:basic_galaxy_properties}. Furthermore, we adopt a flat disk model and added conservative estimates of systematic uncertainty (``error floors") to the data. For the $x$ and $y$
data, we adopt an error floor of 0.015 and 0.030 mas, respectively. For maser velocity, we adopt an error floor of 1.8 km~s$^{-1}$, which is a typical linewidth of a single maser line. These error floors were estimated in a way to make the fitting stable\footnote{For IRAS 08452$-$0011, the fitting becomes unstable when the adopted error floor for x or y is smaller than 0.015 mas.} and allow the reduced $\chi^{2}$ of the fit to be close to one when these values were added in quadrature with the formal uncertainties in the modeling. Finally, we add 311 km~s$^{-1}$
to the observed velocities of all maser spots to reference the maser velocities from the LSR frame to the CMB frame.               
 
Our best fit shows that the position angle P.A. and the inclination
$i$ of the maser disk are P.A. $=$ 207.3$\pm$2.6$^{\circ}$ and $i$ $=$
85.2$\pm$0.3$^{\circ}$, respectively. The reduced $\chi^{2}_{\nu}$ achieved
in our disk modeling is 1.100. Our fit gives a BH mass of $M_{\rm
BH}$ $=$ (3.3$\pm$0.2)$\times$10$^{7}$ $M_{\odot}$ for IRAS 08452$-$0011,
and the dynamical center position obtained from the fit ($x_{0}$
$=$ $-$0.341$\pm$0.008 mas; $y_{0}$ $=$ 0.115$\pm$0.019 mas) is shown
by the yellow star in the left panel of Figure \ref{figure:J0847_plot}.  In the right panel
of Figure \ref{figure:J0847_plot}, we plot the the P$-$V diagram for IRAS 08452$-$0011,
where the dashed line indicates the Keplerian rotation curve calculated
based on the best-fit BH mass. We summarize the best-fit parameters in Table \ref{table:modelparameters}.

Our analysis shown here demonstrates that one is able to apply the
H$_{2}$O megamaser technique to galaxies beyond 200 Mpc for BH mass
measurement with an accuracy of better than $\pm10$\%, well sufficient for
constraining the $M_{\rm BH}-\sigma_{*}$ relation. 
Based on our current analysis of IRAS 08452$-$0011,
which has high-velocity maser flux densities of $\sim40$ mJy
(see Figure \ref{figure:3maser_spectra}), we infer that BH measurements are
feasible up to distances of $\sim$400 Mpc (z $\sim$0.1) with
40 hours observing using the VLBA, augmented with the GBT.  At
such a distance, if the BH mass in a disk maser system is $\sim$10$^{7}$
$M_{\odot}$, the angular size of the maser disk\footnote{Based on the disk
properties of triple-peaked maser systems reported in Kuo et al.
(2011) and Gao et al. (2017), it can be easily shown that all Keplerian
disk maser systems have a characteristic size of $\sim$10$^{5}$ $r_{\rm
s}$, where $r_{\rm s}$ refers to the Schwarzchild radius of the BH.
This suggests that the intrinsic sizes of maser disks would be similar
as long as the BHs have roughly the same masses. For such disks,
if the distance increases by a factor of 2, the angular extent of
the disk would shrink by a factor of $\sim$2.} would shrink
by a factor of $\sim$2 relative to IRAS 08452$-$0011, but the maser
disk can still be well-resolved if the flux densities of maser lines
also increase by a similar factor.

Beyond 400 Mpc (z $\gtrsim$0.1), the applicability of the megamaser
technique for BH mass measurements would be substantially hampered
by sensitivity. With simple
estimation, one can show that such measurement would require the
total maser luminosity $L_{\rm H_{2}O}$ of a disk maser system to
be greater than $\sim$10000 $L_{\odot}$. However, the number of galaxies
with such extreme maser luminosities is negligible in the local Universe (see Table 1 in
Kuo et al. 2018).

Regarding the Hubble constant determination, our result for IRAS
08452$-$0011 suggests that applying the maser technique to galaxies
with distances beyond 200 Mpc would be difficult in the near future
if one aims to determine an $H_{0}$ to better than 10\% accuracy
with a single maser galaxy. Our crude estimate based on our modeling
of the maser disk in NGC 6323 (Kuo et al. 2015) shows that one would
need at least a few hundred hours of observing time to achieve
sufficient maser position accuracy to allow a 10\% $H_{0}$ measurement
with IRAS 08452$-$0011. Nevertheless, the inclusion of the next generation
Very Large Array (ngVLA; McKinnon et al. 2016) in future VLBI observations
is promising to bring about an order of magnitude improvement in
sensitivity which will enable a 1\% $H_{0}$ measurement by efficiently
measuring $\sim$10\% ($\sim$7\%) distances to 100 (50) maser galaxies
with the H$_{2}$O megamaser technique (Braatz et al. 2019). With
the substantially enhanced sensitivity provided by the ngVLA, it
would become possible to make a 10\% $H_{0}$ measurement efficiently
for galaxies beyond 200 Mpc, including IRAS 08452$-$0011. In addition,
it will also facilitate the extension of the application of the megamaser
technique to galaxies beyond 400 Mpc for BH mass measurements because
the necessity of extremely high maser luminosity (i.e. $L_{\rm H_{2}O}$
$>$10000 $L_{\odot}$) will be significantly relaxed.

 \begin{figure*}
\begin{center} 
\vspace*{-2 cm} 
\hspace*{0 cm} 
\includegraphics*[angle=0, scale=0.45]{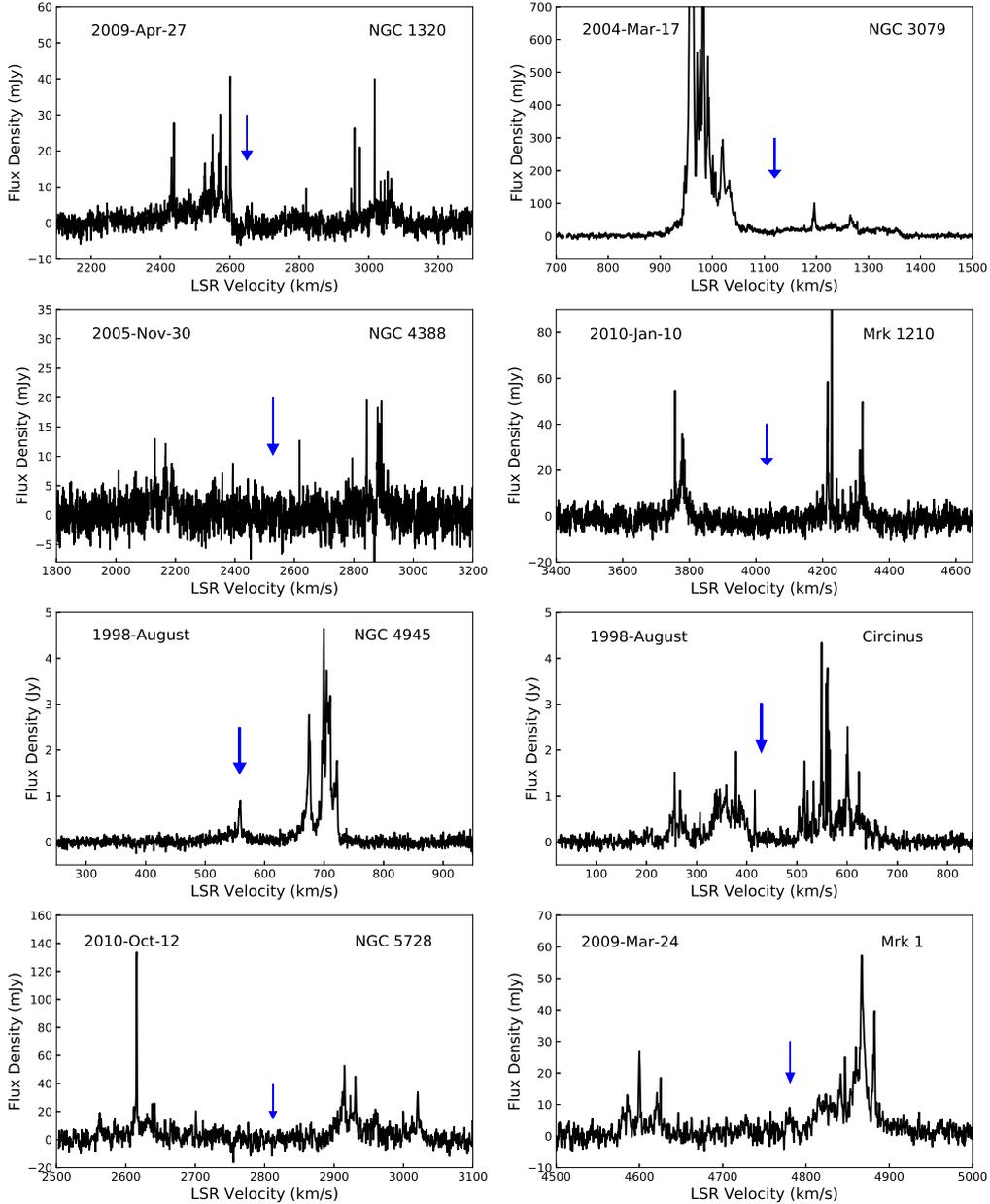} 
\vspace*{-1.5 cm} 
\caption{The spectra of the double-peaked maser systems listed in Table \ref{table:luminosities}. The blue arrow shown in each panel indicates the recessional kinematic
LSR velocity of the galaxy reported in NED. The horizontal axis shows
LSR velocities based on the optical definition. Except for NGC 4945 and Circinus, the single-dish spectra of the double-peaked maser systems were taken with the GBT as part of the MCP maser survey. The spectra for NGC 4945 and Circinus are reproductions of the maser spectra taken with the 64-m Parkes telescope in 1998 August from Braatz et al. (2003). }          
\label{figure:2peakes_maser_spectra}
\end{center} 
\end{figure*}

\begin{figure*}
\begin{center} 
\vspace*{0 cm} 
\hspace*{-2 cm} 
\includegraphics[angle=0, scale=0.42]{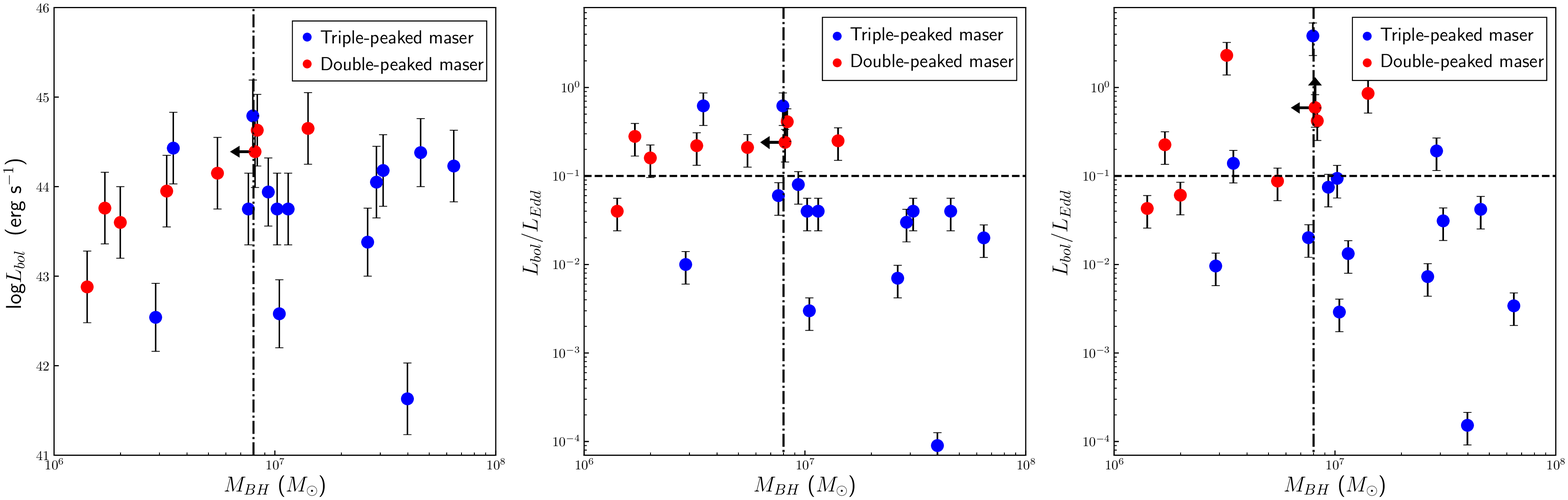} 
\vspace*{0.0 cm} 
\caption{\label{fig:image} {\bf Left panel:} The AGN bolometric luminosity$-$BH
mass ($L_{\rm bol}$$-$$M_{\rm BH}$) diagram. The blue and red dots
indicate the triple-peaked and double-peaked H$_{2}$O megamaser systems,
respectively. The dot-dashed line indicates a BH mass of 8$\times$10$^{6}$
$M_{\odot}$. $L_{\rm bol}$ shown here are mainly inferred from intrinsic X-ray luminosities. When the X-ray measurement is not available for a galaxy, we use [OIII] luminosity to infer $L_{\rm bol}$. The error bars shown in the plot reflect the mean uncertainty of the $L_{\rm bol}$ estimation. {\bf Middle panel:} The Eddington ratio$-$BH mass ($\lambda_{\rm
Edd}$$-$$M_{\rm BH}$) diagram. The horizontal dashed line indicates
an Eddington ratio $\lambda_{\rm Edd}$ $\equiv$ $L_{\rm bol}/L_{\rm
Edd}$ of 0.1 whereas the vertical dot-dashed line shows a BH mass
of 8$\times$10$^{6}$ $M_{\odot}$.  $\lambda_{\rm Edd}$ are evaluated from $L_{\rm bol}$ shown in the left panel. The double-peaked maser systems
tend to have higher $\lambda_{\rm Edd}$ and lower $M_{\rm BH}$ while
the triple-peaked megamasers show an opposite trend.  {\bf Right panel:} The $\lambda_{\rm
Edd}$$-$$M_{\rm BH}$ diagram with $\lambda_{\rm
Edd}$ evaluated from $L_{\rm bol}$ inferred from [OIII] luminosity. This plot does not include NGC 4945, which is the only source in Table \ref{table:luminosities} that does not have [OIII] luminosity available. This plot shows that the statistical trends remain the same for the double-peaked and triple-peaked maser systems when we adopt a different tracer for AGN bolometric luminosity.  }   
\label{figure:lambda_mass_diagram}                
\end{center} 
\end{figure*} 
 
 \section{Discussion} 

\subsection{The Association between Perturbed Disk and AGN Winds} 

In the previous sections, we see that the masers in the double-peaked
H$_{2}$O megamaser systems NGC 5728 and Mrk 1 have substantially
different maser distributions and kinematics in comparison with those
of thin, edge-on Keplerian maser disks such as NGC 4258. While the
spatial sizes of these two double-peaked maser systems are similar
to the typical sizes of triple-peaked maser disks (Gao et al. 2017),
their spatial distributions do not directly reveal that the H$_{2}$O
masers reside in thin, unperturbed, gas disks. In addition,
their kinematics are unusual and do not trace Keplerian rotation curves 
that would allow reliable fitting of BH masses.
 
These differences argue against the double-peaked megamasers in NGC
5728 and Mrk 1 being simply triple-peaked disk masers in which the
systemic maser lines are significantly weaker than the high-velocity
masers.  To investigate whether this is a general property of double-peaked
H$_{2}$O megamasers, we collect from the literature all megamaser systems whose 
spectra are dominated by two distinct line complexes and have the
VLBI maser maps and kinematic measurements that allow for
estimates of the enclosed mass. The
references from which we obtain the maser maps and BH/enclosed mass
estimates are shown in Table \ref{table:luminosities}. The spectra of these double-peaked megamasers  
are shown in Figure \ref{figure:2peakes_maser_spectra}.

From Figure \ref{figure:2peakes_maser_spectra}, it can be seen that all of the double-peaked megamasers except for NGC 4945 have two maser line complexes that are redshifted and blueshifted with respect to the recession velocities ($V_{\rm sys}$) of the galaxies. For all of these systems, there are occasionally weak or narrow (i.e. linewidth $\lesssim$ 5 km~s$^{-1}$) maser lines arising in between the two maser complexes in some epochs and these lines often lie within 100 km~s$^{-1}$ from $V_{\rm sys}$ of these sources. While these maser features could be the systemic maser lines as found in triple-peaked maser systems, they are also likely to be maser lines arising at the outskirts of the redshifted/blueshifted complexes. Based on their locations in the single-dish spectra alone, these tentative systemic maser lines cannot be easily and clearly distinguished from high velocity maser features. For NGC 4945, the blueshifted maser complex is missing and there is a line complex whose peak velocity well agrees with $V_{\rm sys}$ of the galaxy. Thus, this is likely to be the systemic maser complex as seen in the spectra of the triple-peaked systems.

From the VLBI maps of the double-peaked systems, we note that their maser distributions and kinematics are
similar to those of NGC 5728 and Mrk 1, with
some, or all, of the maser features tracing a roughly linear distribution
on the sky that suggests a disk.  However, the degree of scatter seen in 
the maser distributions often allows for alternative interpretations, 
such as the presence of outflows. 
Moreover, no P$-$V diagram can be described by a ``clean'',
smoothly varying rotation curve. Even if the maser velocity falls
as a function of radius, the intrinsic scatter of maser (e.g. Mrk 1210;
see Zhao et al. 2018) seen in the rotation curve is often substantially greater than the measurement uncertainties (i.e. $\sim$1$-$2 km~s$^{-1}$).

From the above comparison, we see evidence that the H$_{2}$O masers
in the double-peaked systems tend to reside in dynamically perturbed
gas disks. It is likely that non-gravitational forces such as AGN
winds have a stronger impact on the masing gas in the double-peaked
megamasers than in the triple-peaked maser systems, leading to 
different maser distributions and kinematics.

Indeed,  when we look at the multi-wavelength imaging of the double-peaked
systems, we can always see evidence for AGN winds or jets
in comparison with the triple-peaked systems. For example, Circinus shows a prominent wind traced by H$_{2}$O maser emission at $\sim$1 pc scale (Greenhill et al. 2003).
For NGC 1320 (Mrk 607),  gas kinematics at $\sim$300 pc scale show counter rotation with respect to stars in the galactic disk, suggesting that the gas is tracing an equatorial outflow (Freitas et al. 2018). NGC 4945 shows kpc-scale nuclear outflow cones suggesting the presence of a starburst-driven wind (Heckman 2003). NGC 3079 has a bipolar super-wind that inflates a kpc-scale superbubble (Duric \& Seaquist 1988; Veilleux et al. 1994; Cecil et al. 2002) and show evidence for a wide-angle AGN-driven outflow on parsec-scales (Kondratko et al. 2005). Prominent
jets which have physical extents ranging from a few to $\sim$30 parcsec
can be seen in the double-peaked maser systems Mrk 1 and
Mrk 1210 (Kamali et al. 2019). In NGC 5728,
one can even see a kpc-scale collimated radio jet propagating in
the direction that aligns with the ionization cone (Durr$\acute{\rm
e}$ \& Mould 2018) that traces a bipolar outflow. Finally, in NGC 4388, the [OIII] image shows an ionization cone at 100$-$400 pc scale (Greene et al. 2014), and the radio emission morphology suggests a collimated AGN-driven outflow (Stoklasov$\acute{\rm a}$ et al. 2009; Stone et al. 1988; Falcke et al. 1998). 

On the other hand, among the 14 triple-peaked maser systems listed in
Table \ref{table:luminosities}, only 4 galaxies (NGC 4258, NGC 1068, NGC 3393, and IC 2560) show clear evidence for jets :  NGC 4258 and NGC 1068 show both parsec and kpc scale jets (Herrnstein et al. 1999; Gallimore et al. 2001, 2004); NGC 3393 has a kpc scale double-sided jet (Schmitt et al. 2001); for IC 2560, a jet-like continuum is found at parsec scale (Yamauchi et al. 2012). For the rest of the triple-peaked sources, evidence for an outflow or jet is not obvious. While there are no observations that could determine whether jets/outflows exist in NGC 1194, NGC 5495, NGC 5765b, and UGC 6093, Kamali et al. (2019) conducted radio jet imaging toward some of these maser systems and found that radio emission at the milliarcsecond scale is either absent (e.g. UGC 3789, NGC 6323, NGC 6264) or shows a structure that does not
necessarily reflect the presence of a jet (e.g. NGC 2273, NGC 2960). 

This suggests that the two types of megamasers may arise from dynamically different environments, with the double-peaked megamasers possibly residing in AGN where winds/outflows are prominent on a $\sim$1 pc scale (i.e. the typical size of maser disks) and have greater impact on the dynamics of the masing gas. It is likely that wind disturbances substantially reduce the coherence (amplification) path lengths of maser emissions, making strong maser emission harder to occur. As a result, the three maser complexes in a pristine triple-peaked maser disk may not be always present when a wind-perturbed maser source is detected. Depending on the level of the wind disturbance, one or more maser line complexes of a disk maser system could be missing. In the well-studied maser source Circinus, the missing component is the systemic maser complex whereas in the case of NGC 4945, the blueshifted maser features are not clearly present in the spectrum (see Figure \ref{figure:2peakes_maser_spectra}).

\begin{table*}
	\centering
	\caption{The AGN properties of the Triple-peaked and Double-peaked
H$_{2}$O Megamasers}
	\label{table:luminosities} 
	\begin{tabular}{lccllcccrrrr} % four columns, alignment for each
		\hline\hline
    & Disk & log$M_{\rm BH}$  & log$L^{\rm int}_{2-10}$   & log$L_{\rm [OIII]}$
& log$L_{\rm bol,x}$   & log$L_{\rm bol,[OIII]}$  & log$L_{\rm Edd}$ & $\lambda_{\rm Edd}$ & Ref.$^{\rm a}$
  & Ref.$^{b}$ & Ref.$^{b}$ \\                          
Name   & Type   & ($M_{\odot}$)   & (erg~s$^{-1}$) & (erg~s$^{-1}$)   & (erg~s$^{-1}$) & (erg~s$^{-1}$)    & (erg~s$^{-1}$)
&   & ($M_{\rm BH}$)  & ($L_{X}$)  & ($L_{\rm [OIII]}$) \\
	\hline
NGC 1068 & III & 6.90 & 43.34 & 42.8 & 44.79 & 45.58 & 45.00 & 0.62 & 11 & 13 & 23 \\     
NGC 1194  & III  &  7.81 & 42.78 & 39.9$^{\rm c}$  & 44.23 &  43.44 & 45.90 & 0.02 & 1
& 13  & 17   \\                                                               
NGC 2273  & III  & 6.88 &  42.30 & 40.5 & 43.75 & 43.28 & 44.96 & 0.06 & 1
& 14 & 17 \\                                                                 
NGC 2960  & III  & 7.06 &  42.30 & 40.5 & 43.75 & 43.28 & 45.15 & 0.04 & 1
& 13 & 17\\                                                                 
NGC 3393 &  III & 7.49 & 42.73 & 41.3 & 44.18 & 44.08 & 45.58 & 0.04 & 2 & 15 & 17\\ 
NGC 4258 &  III & 7.60 &  40.63 & 39.1 & 41.63 & 41.88 & 45.69 & 0.00009 &
3 & 15 & 23 \\                                                               
NGC 5495  & III  &  7.02 & --- & 39.8 & --- & 42.58 & 45.11 & 0.003 & 4
& --- & 4\\                                                                
NGC 5765b  & III  & 7.66 &  --- & 41.6 & --- & 44.38 & 45.76 & 0.04 & 5
& --- & 16\\                                                                 
NGC 6323 & III  & 6.97 &  --- & 40.4$^{\rm c}$ & --- & 43.94 & 45.06 & 0.08 & 1 & --- & 17 \\ 
NGC 6264 & III & 7.46 & 42.60 & 41.3$^{\rm c}$ & 44.05 & 44.84 & 45.55 & 0.03 & 1 & 18 & 17\\ 
UGC 3789 & III & 7.01 &  42.30 & 41.3 & 43.75 & 44.08 & 45.10 & 0.04 & 1 & 18 & 17 \\ 
J0437$+$2456 & III &  6.46 & --- & 39.0$^{\rm c}$ & --- & 42.54 & 44.55 & 0.01 &
4 & --- & 4 \\    
IC 2560 & III & 6.54 & 42.98 & 41.0 & 44.43 & 43.78 & 44.64 &  0.62 & 12 & 13 & 17 \\                                                      
UGC 6093  & III  &  7.42 & --- & 40.6 & --- & 43.38 & 45.51 & 0.007 & 6
& ---  & 16 \\                                                               
NGC 1320  & II  & 6.74 &  42.70 & 41.0 & 44.15 & 43.78 & 44.83 & 0.21 & 4
& 19 & 22 \\                                                                 
NGC 3079 & II & 6.30 & 42.15 & 40.4 & 43.60 & 43.18  & 44.39 & 0.16 & 7 & 20 & 22\\ 
NGC 4388 & II & 6.92 & 43.18 & 41.9 & 44.63 & 44.70 & 45.01 & 0.42 & 1 & 15 & 24 \\ 
NGC 4945 & II & 6.15 &  41.43 & --- & 42.88 & --- & 44.23 & 0.04 & 8 & 15 & ---\\ 
NGC 5728  & II  &  $<$6.91 & 42.94 & 42.0 & 44.39 & 44.78 & $<$45.01 & $>$0.24 & 9
& 15  & 22 \\                                                               
Mrk 1 & II  & 6.51 &  42.50 & 41.4$^{\rm c}$ & 43.95 & 44.97 & 44.60 & 0.22 & 9 & 21 & 24\\ 
Mrk 1210 & II & 7.15 & 43.20 & 42.4 & 44.65 & 45.18 & 45.24 & 0.25 & 6 & 15  & 22\\ 
Circinus & II & 6.23 &  42.32 & 40.9 & 43.76 & 43.68 & 44.32 & 0.28 & 10 & 15 & 22 \\
              \hline     
 \multicolumn{12}{|p{16.5 cm}|}{Note. Col(1): Name of the maser galaxy; Col(2): The maser
disk type. Type II \& III refer to the double-peaked and triple-peaked
megamasers, respectively; Col(3): The BH mass in units of solar mass;
Col(4); The absorption-corrected intrinsic 2-10 keV X-ray luminosity;
Col(5): The [OIII] luminosity $L_{\rm [OIII]}$ (erg~s$^{-1}$). Except for NGC 1194, NGC 6323, NGC 6264, J0437$+$2456, and Mrk 1, the $L_{\rm [OIII]}$ measurements include internal reddening corrections; Col(6): The bolometric luminosity of the AGN estimated
from $L^{\rm int}_{2-10}$. For those galaxies with X-ray measurements except for NGC 4258, given
their nature of being either highly obscured (i.e. the obscuring
column density $N_{\rm H}$ $\ge$ 10$^{23.5}$ cm$^{-2}$) or Compton-thick
($N_{\rm H}$ $\ge$ 10$^{24}$ cm$^{-2}$), we obtained $L_{\rm bol}$
by applying a bolometric correction factor $\kappa_{bol}$ of 28 (Brightman
et al. 2017) to $L^{\rm int}_{2-10}$. For NGC 4258, which hosts a
Compton-thin nucleus, we adopt a $\kappa_{bol}$ of 10 (Lusso et al.
2012) for the bolometric correction; Col(7): The bolometric luminosity of the AGN estimated
from $L_{\rm [OIII]}$ (erg~s$^{-1}$). The adopted bolometric correction factors are 600 (Heckman \& Best 2014) and 3500 (Heckman et al. 2004) for $L_{\rm
[OIII]}$ with and without internal reddening corrections, respectively; Col(8): The Eddington Luminosity (erg~s$^{-1}$)
calculated using the BH mass shown in column (3); Col(9): The Eddington
ratio $\lambda_{\rm Edd}$ $\equiv$ $L_{\rm bol}/L_{\rm Edd}$, with $L_{\rm bol}$ primarily derived from $L^{\rm int}_{2-10}$. When X-ray flux measurements are not available, we adopt $L_{\rm bol}$ estimated from [OIII] luminosity to evaluate $\lambda_{\rm Edd}$; Col(10):
Reference for VLBI imaging and the maser-based BH mass measurement;
Col(11): Reference for the measurement of $L^{\rm int}_{2-10}$; Col(12): Reference for the measurement of $L_{\rm [OIII]}$. }\\ 
\multicolumn{12}{|p{16.5 cm}|}{$\rm ^a$ {\bf References for the BH mass $M_{\rm BH}$ measurement}
: 1. Kuo et al. (2011); 2. Kondratko et al. (2008); 3. Humphyreys
et al. (2013); 4. Gao et al. (2017); 5. Gao et al. (2016); 6. Zhao
et al. (2018); 7. Kondratko, Greenhill, \& Moran (2005); 8. Greenhill
et al. (1997); 9. this paper; 10. Greenhill et al. (2003).}\\  
\multicolumn{12}{|p{16.5 cm}|}{$\rm ^b$ {\bf References for $L^{\rm int}_{2-10}$ and $L_{\rm
[OIII]}$} : 13. Masini et al. (2016); 14. Awaki et al. (2009);
15. Ricci et al. (2017); 16. MPA-JHU emission line analysis for the
SDSS data (${\rm http://www.sdss3.org/dr10/spectro/galaxy_mpajhu.php}$);
17. Greene et al. (2010); 18. Castangia et al. (2013); 19. Brightman
\& Nadra (2011); 20. Panessa et al. (2006); 21. Singh et al. (2011); 22. Shu et al. (2007); 23. Ho et al. (1997); 24.  Zhu et al. (2011)}\\  	
\multicolumn{12}{|p{16.5 cm}|}{$\rm ^c$ For these sources, flux measurements of either
$H_{\alpha}$ or $H_{\beta}$ are not available for calculating the
Balmer decrement, no internal reddening correction is applied to observed $L_{\rm [OIII]}$}\\  		 
	 
	  \end{tabular}

\end{table*}

\subsection{The $\lambda_{\rm Edd}$ $-$ $M_{\rm BH}$ Diagram} 

To evaluate the relative impacts of AGN winds on the double-peaked
and triple-peaked H$_{2}$O megamaser disks, it is helpful to compare
the AGN bolometric
luminosity ($L_{\rm bol}$)  or the Eddington ratio ($\lambda_{\rm Edd}$) between these two types of maser systems.
In recent studies of AGN feedback, it is found that properties of
AGN winds are well-correlated with either the AGN power or accretion
efficiency. Greene (2006), Woo et al. (2016) and Kang et al. (2017) show that the
fraction of AGN that reveal clear signatures of ionized gas outflows
(i.e. the non-gravitational components of the [OIII] linewidths)
increase rapidly as $L_{\rm bol}$ and $\lambda_{\rm Edd}$ go up.
In addition, Fiore et al. (2017) demonstrate that the mass outflow
rate, wind kinetic power and the maximum wind velocity of both ionized
and molecular gas show strong positive correlations with the AGN
power. Furthermore, Cicone et al. (2014) and Combes et al. (2014)
indicate that the outflow momentum of a molecular wind also increases
with the AGN luminosity. Finally, from the theoretical point of view,
Giustini \& Proga (2019) show that in AGN with $\lambda_{\rm Edd}$
is $\gtrsim$0.1$-$0.25, the radiation pressure becomes large enough
to allow for the production of strong, persistent disk winds whereas
for AGN with $\lambda_{\rm Edd}$ $\lesssim$0.1, only a relatively
weak or ``failed'' disk wind could occur.  Therefore,
if the double-peaked megamasers are systems more disturbed by AGN
winds, one might expect to see a correlation in these maser systems with
$L_{\rm bol}$ or $\lambda_{\rm Edd}$.  We now investigate this possibility.

In the left panel of Figure \ref{figure:lambda_mass_diagram}, we plot $L_{\rm bol}$ as a function of $M_{\rm BH}$ for the twenty-two
H$_{2}$O megamaser systems listed in Table \ref{table:luminosities}, with the triple-peaked
and double-peaked megamasers color-coded in blue and red, respectively. The AGN bolometric luminosities $L_{\rm bol}$ of these maser systems
are mainly derived using the absorption-corrected intrinsic 2-10
keV X-ray luminosities of the AGN. For those galaxies which have X-ray measurements available (except for NGC 4258), given
their nature of being either highly obscured (i.e. the obscuring
column density is $N_{\rm H}$ $\ge$ 10$^{23.5}$ cm$^{-2}$) or Compton-thick
($N_{\rm H}$ $\ge$ 10$^{24}$ cm$^{-2}$), we obtained $L_{\rm bol}$
by applying a bolometric correction factor $\kappa_{bol}$ of 28 (Brightman
et al. 2017) to $L^{\rm int}_{2-10}$. For NGC 4258, which hosts a
Compton-thin nucleus, we adopt a $\kappa_{bol}$ of 10 (Lusso et al.
2012) for the bolometric correction. When reliable X-ray luminosities
are not available for a particular source, we use the [OIII] luminosity ($L_{\rm [OIII]}$)
to infer $L_{\rm bol}$. When estimating $L_{\rm bol}$ from $L_{\rm [OIII]}$, we adopt bolometric correction factors of 600 (Heckman \& Best 2014) and 3500 (Heckman et al. 2004) for $L_{\rm
[OIII]}$ with and without internal reddening corrections, respectively.

\begin{figure} 
\begin{center} 
\vspace*{0 cm} 
\hspace*{0 cm} 
\includegraphics*[angle=0, scale=0.5]{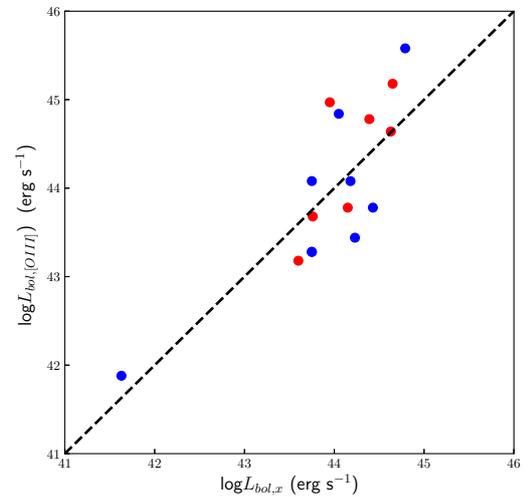} 
\vspace*{0 cm} 
\caption{Bolometric luminosities estimated from absorption corrected 2-10 keV X-ray luminosity (horizontal axis) and [OIII] luminosity (vertical axis). The red and blue dots incidate the double-peaked and triple-peaked maser systems, respectively. The scatter of the $L_{\rm bol,x}$$-$$L_{\rm bol,[OIII]}$ relation is 0.54 dex.}          
\label{figure:luminosity_consistency}
\end{center} 
\end{figure}

To assess the uncertainty of $L_{\rm bol}$ obtained from X-ray measurements ($L_{\rm bol,x}$), we compare $L_{\rm bol,x}$ with the bolometric luminosities estimated from [OIII] luminosities ($L_{\rm bol,[OIII]}$), which are available for all but one galaxy (NGC 4945) listed in Table \ref{table:luminosities}. This comparison is shown in Figure \ref{figure:luminosity_consistency}. It can be seen from this figure that $L_{\rm bol}$ inferrred from $L_{2-10}^{int}$ and $L_{\rm [OIII]}$ are consistent. The mean and standard deviation of $\Delta L_{\rm bol}$ $\equiv$ $L_{\rm bol,x}$ $-$ $L_{\rm bol,[OIII]}$ are $-$0.05 and 0.54 dex, respectively. Assuming the uncertainty of the bolometric correction ($\sim$0.38 dex; Heckman et al. 2004) dominates the error budget for $L_{\rm bol,[OIII]}$, the standard deviation of $\Delta L_{\rm bol}$ suggests that the mean uncertainty of $L_{\rm bol,x}$ is 0.42 dex, comparable to the uncertainty of $L_{\rm bol,[OIII]}$.

In the left panel of the Figure \ref{figure:lambda_mass_diagram}, one can see that there is no strong correlation between  
AGN luminosities and the two types of maser systems.
On the other hand, the $L_{\rm bol}$$-$$M_{\rm BH}$ diagram
reveals that triple-peaked maser disks tend to have greater BH masses than double-peaked maser systems. Relative to a reference BH mass of
$M_{\rm BH}$ $=$ 8$\times$10$^{6}$ $M_{\odot}$, which is chosen to clearly separate the $M_{\rm BH}$ distributions of the two types of megamasers, 71\% (75\%) of triple-peaked (double-peaked) megamasers have BH masses greater (smaller) than the reference value.
The systematically lower BH masses of the double-peaked megamasers
could suggest that the BH mass plays a role in determining
the type of maser system. Alternatively, the lower BH masses could also result from stronger winds in these systems 
which perturb the maser kinematics and lead to underestimations ($\Delta M_{\rm turb}$)
of the BH masses. However, we found this latter interpretation untenable. Our conservative estimates of $\Delta M_{\rm turb}$ for double-peaked systems show that the ratios between $\Delta M_{\rm turb}$ and $M_{\rm BH}$ range from 1$-$34\% (see Appendix A), suggesting that the effect of AGN wind perturbation is not strong enough to account for the systematic BH mass difference between the two types of maser systems.
  
Although the double-peaked maser systems do not show a different $L_{\rm bol}$ distribution with respect to triple-peaked megamasers, we see evidence that these two types of maser systems have systematically different $\lambda_{\rm Edd}$. In the middle panel of Figure \ref{figure:lambda_mass_diagram}, it can be seen that the double-peaked megamasers tend to have Eddington ratios\footnote{The Eddington ratios $\lambda_{\rm Edd}$ discussed here are evaluated with $L_{\rm bol}$ primarily inferred from the AGN X-ray luminosities. When reliable X-ray luminosity is not available for a particular source, we evaluate $\lambda_{\rm Edd}$ based on $L_{\rm bol}$ inferred from the [OIII] luminosity.  } greater than the theoretical threshold above which disk winds become strong and persistent (i.e. $\lambda_{\rm
Edd}$ $\simeq$ 0.1; Giustini \& Proga 2019) whereas the triple-peaked maser systems show an opposite
trend. To test whether this statistical tendency is robust against the methods of estimating $L_{\rm bol}$, we re-evaluate $\lambda_{\rm Edd}$ by inferring $L_{\rm bol}$ purely based on [OIII] luminosities, and we show the resulting $\lambda_{\rm Edd}$$-$$M_{\rm BH}$ diagram in the right panel of Figure \ref{figure:lambda_mass_diagram}. One can see that double-peaked megamasers on average still have higher Eddington ratios than triple-peaked disk masers when $\lambda_{\rm Edd}$ are inferred from a different tracer of $L_{\rm bol}$, and $\lambda_{\rm Edd}$ remain $\gtrsim$ 0.1 for the majority of the double-peaked systems, consistent with the result with $L_{\rm bol}$ inferred from X-ray luminosities.

If AGN winds indeed become more prominent in higher $\lambda_{\rm
Edd}$ systems as suggested by Woo et al. (2016), Kang et al. (2017),
and Giustini \& Proga (2019), this could explain the differences.
In this interpretation, the double-peaked
maser systems are gas disks residing in AGN with $\lambda_{\rm Edd}$
$\gtrsim$ 0.1 in which disk winds driven by radiation pressure become
strong and persistent (Giustini \& Proga 2019). It is in these
AGN that a gas disk is more likely to get disturbed by a wind,
and masers also have a greater chance to appear in an outflow. On the other hand, the Keplerian triple-peaked maser disks
prefer lower Eddington ratio AGNs (i.e. $\lambda_{\rm Edd}$ $<$ 0.1),
in which winds/outflows are weaker and allow the dynamics
to be dominated by the central BH.

An alternative way to explain the observed trends seen in the $\lambda_{\rm Edd}$$-$$M_{\rm BH}$ diagram is that the triple-peaked maser disks reside in optically thick, geometrically thin disks whereas masers in the double-peaked systems tend to reside in geometrically thick {\it slim} accretion disks (Czerny 2020). Here, the slim accretion disk models describe an optically thick, geometrically not very thin accretion flow around a black hole in high Eddington ratio AGN where the radiation pressure is enhanced and drives/supports the thickness of the accretion disk.
 
 As discussed in Koratkar \& Blaes (1999), geometrically thin disk models are not consistent with accretion rates above $\eta_{\rm Edd}$ $\gtrsim$0.2$-$0.3, where a thick/slim disk geometry and advection/convection must be involved (e.g. Sadowski et al. 2011). This is perhaps one of the reasons why nearly all of the geometrically thin, Keplerian triple-peaked maser disks listed in Table \ref{table:luminosities} have $\eta_{\rm Edd}$ $<$ 0.2. Indeed, when we observe the maser distributions in all of the double-peaked maser systems, in no cases the thickness of the disk can be ignored in comparison with the disk radius, supporting the view that these masers may reside in geometrically thick disks which occur in high Eddington ratio AGN. Nonetheless, we note that the high $\lambda_{\rm Edd}$ thick disk scenario alone may not be able to fully explain the maser kinematics of the double-peaked maser systems. This is because the gas motions affected by convection/advection will mainly occur in the radial and vertical direction of the disk and may not significantly affect the gas velocity along the line-of-sight for the gas residing close to the mid-line of the disk (i.e. the high-velocity masers). Thus, this scenario could not easily account for the significant deviations from Keplerian rotation seen in the P$-$V diagrams of all double-peaked maser systems. Non-gravitational forces such as winds and outflows would be still needed to explain these deviations.  
 
 Finally, one should be aware that our main interpretation for double-peaked maser systems may not be universal and there could be exceptions. If a classic, low $\lambda_{\rm Edd}$ triple-peaked maser system such as NGC 4258 (Herrnstein et al. 1999) has a high velocity complex that is much weaker than the other two complexes (possibly due to significant disk warping), the system would appear to be double-peaked if the sensitivity is not sufficient. In this case, the absence of a maser line complex would be due to a warp and not due to an AGN wind. Therefore, one can expect that there would be exceptions in the $\lambda_{\rm Edd}$$-$$M_{\rm BH}$ diagram especially if there are the double-peaked maser systems that have $\lambda_{\rm Edd}$ $\ll$ 1 (i.e. less prominent winds). For such exceptions, single-dish spectra alone may not be sufficient to characterize the nature of these maser systems. One would need VLBI images to help determining whether these are wind-perturbed systems or triple-peaked warped maser disks which display only two maser complexes due to insufficient sensitivity.

\section{Conclusions}

In this work, we classify H$_{2}$O megamaser galaxies into triple-peaked and
double-peaked  maser systems. 
Among the three maser galaxies we study
in this paper, IRAS 08452$-$0011 belongs to the triple-peaked megamaser
whereas NGC 5728 and Mrk 1 belong to the less explored double-peaked
systems. Our main conclusions are summarized as follows:              
 
\begin{itemize} 
\item[1.] The maser distribution and kinematics in NGC 5728 may result
from masing gas tracing a rotating magnetocentrifugal wind. For Mrk 1,  the blueshifted masers reside in a disk while the majority of the redshifted masers may follow a wind.
The BH mass in NGC 5728 is shown to be smaller than 8.2$\times$10$^{6}$ $M_{\odot}$. The rotation curve fitting for the blueshifted masers in Mrk 1 gives a BH mass of (3.2$\pm$0.5)$\times$10$^{6}$ $M_{\odot}$. 
 
\item[2.] The maser distribution and kinematics in 
IRAS 08452$-$0011 are consistent with a gas disk
with Keplerian rotation.  Modeling of the disk in three dimensions
gives a BH mass of $M_{\rm BH}$ $=$ (3.3$\pm$0.2)$\times$10$^{7}$
$M_{\odot}$. This measurement demonstrates that the H$_{2}$O megamaser
technique can be applied to galaxies beyond 200 Mpc for BH mass measurements.

\item[3.] Disturbed maser distributions and kinematics are a ubiquitous
feature of all double-peaked maser systems, in which signatures of
AGN outflows or winds appear to be more prominent than in triple-peaked
megamasers, suggesting that the disturbed maser distribution and
kinematics are associated with AGN winds which occur on a $\sim$1
pc scale.                                                             
 
\item[4.] The double-peaked and triple-peaked maser systems show
distinctly different distributions in the $\lambda_{\rm Edd}$$-$$M_{\rm
BH}$ diagram.  The triple-peaked masers tend to have $\lambda_{\rm
Edd}$ $<$ 0.1 while the double-peaked sources show an opposite trend.
This supports the picture that double-peaked systems tend to reside
in an environment where the circumnuclear gas has a higher chance
to get disturbed dynamically because of a more prominent wind. On
the other hand, in a triple-peaked maser system, the AGN wind becomes
less prominent, allowing the disk to be dominated
by the gravity of the BH and enabling the gas to follow Keplerian rotation.\\   
 
\end{itemize}

\section*{Acknowledgements}

The National Radio Astronomy Observatory is a 
facility of the National Science Foundation operated under cooperative 
agreement by Associated Universities, Inc. This publication is supported
by the Ministry of Science and Technology, R.O.C. under the project 108-2112-M-110-002.
This                                                                    
research has made use of NASA's Astrophysics Data System Bibliographic 
Services, and the NASA/IPAC Extragalactic Database (NED) which is 
operated by the Jet Propulsion Laboratory, California Institute of 
Technology, under contract with the National Aeronautics and Space 
Administration.

\appendix

\section{Estimates of the systematic uncertainties in the BH mass measurements caused by wind disturbances}

To assess the magnitude of BH mass underestimation for a maser disk perturbed by a wind, we assume that the disturbance from the wind injects kinetic energy into the maser system and drives/ enhances turbulence in the maser disk.
When the gas pressure caused from turbulence is not negligible, the orbital velocity of a maser spot in the disk can be expressed as
\begin{equation}
v_{\rm orb} = \sqrt{ {GM_{\rm BH} \over r} + \frac{r}{\rho}{dP_{\rm gas} \over dr} }~,
\end{equation}
where $r$ is the radial distance of the maser spot from the dynamical center, $\rho$ is the local volume density, and $P_{\rm gas}$ is the gas pressure at radius $r$ (Haworth et al. 2018).
In a turbulent disk, $P_{\rm gas}$ includes both thermal and turbulent components. So, $P_{\rm gas}$ $=$ $\rho c_{s}^{2}$ $+$ $\rho <v_{t}^{2}>$, where $c_{s}$ is the isothermal sound speed and $<v_{t}^{2}>$ is the squared turbulent velocity dispersion (Montesinos Armijo \& de Freitas Pacheco 2011). Assuming that the turbulent pressure dominates over the thermal pressure in a wind-perturbed maser disk, $P_{\rm gas}$ $=$ $\rho <v_{t}^{2}>$.

\begin{table}
	\centering
	\caption{BH mass underestimation caused by wind disturbance }
	\label{table:mass_underestimation} 
	\begin{tabular}{ lcccc} % four columns, alignment for each
		\hline\hline
  & $r_{\rm in}$  & $v_{\rm turb}$  & $\Delta M_{\rm turb}/M_{\rm enc}$   & $\Delta M_{\rm dist}/M_{\rm enc}$  \\                          
Name    & (pc)  & (km~s$^{-1}$)  & (\%) & (\%)\\
	\hline
NGC 1320  & 0.07  & 40 &  1 & 7.0 \\                                                            
NGC 3079 & 0.40 & 60 & 34 & 6.8 \\ 
NGC 4388 & 0.24 & 10 & 0.1 & 7.2  \\ 
NGC 4945 & 0.16 & 60 &  24 & 7.3 \\                                                               
Mrk 1 & 0.11  & 30 &  2 & 7.0\\ 
Mrk 1210 & 0.26 & 30 & 1 & 7.0 \\ 
Circinus & 0.11 & 30 &  3 & 7.3 \\
              \hline     
 \multicolumn{5}{|p{8 cm}|}{Note. Col(1): Name of the maser galaxy; Col(2): The inner radius of the maser disk ($r_{\rm in}$); Col(3): The conservative estimate of the turbulent velocity dispersion ($v_{\rm turb}$) inferred from the position-velocity diagram of the maser system; Col(4): The ratio between BH mass underestimation ($\Delta M_{\rm turb}$) caused by wind disturbance and the BH mass estimate ($M_{\rm enc}$) of the maser system reported in Table \ref{table:luminosities}. $\Delta M_{\rm turb}$ is evaluated with $r_{\rm in}$ and $v_{\rm turb}$ shown in column (2) \& (3); The ratio between $\Delta M_{\rm dist}$ and $M_{\rm enc}$, where $\Delta M_{\rm dist}$ refers to the BH mass error caused by uncertainty of galaxy distance reported in NED.  }\\              
	\end{tabular}
\end{table}

Based on the standard theory of accretion disks, $\rho(r)$ $=$ $\Sigma(r)$/[$\sqrt{2\pi}$$H(r)$], where $\Sigma(r)$ and $H(r)$ are disk surface density and scale height at radius $r$, respectively (Neufeld \& Maloney 1995). For a turbulence dominated disk, the disk height is characterized by $H(r)$ $=$  $v_{\rm turb}$/$\Omega$ (Vollmer \& Davies 2013), where $v_{\rm turb}$ is turbulence velocity dispersion $v_{\rm turb}$ $\equiv$ $<v_{t}^{2}>^{1/2}$ and $\Omega$ is the angular velocity of the disk. For the double-peaked maser systems listed in Table \ref{table:luminosities} except for NGC 3079, $\Omega$ can be approximated by Keplerian angular velocity $\Omega_{\rm K}$ $=$ $\sqrt{GM_{\rm BH}/r^{3}}$. For NGC 3079 (Kondratko et al. 2005), which has a nearly flat rotation curve, $\Omega$ $\propto$ 1/$r$. Assuming that maser disk surface density follows $\Sigma(r)$ $\propto$ $r^{-1}$ (Hur$\acute{\rm e}$ et al. 2011), one can infer that $\rho$ $\propto$ $r^{-\eta}$, where $\eta$ $=$ 2 for NGC 3079 and $\eta$ $=$ 5/2 for the rest of double-peaked disk masers. In addition, one can easily verify that 
\begin{equation}
\frac{r}{\rho}{dP_{\rm gas} \over dr} = -\eta v_{\rm turb}^{2}~.
\end{equation}
if we assume that $v_{\rm turb}$ is constant across the masing region of the disk. With this, Equation (1) can be re-written as
\begin{equation}
v_{\rm orb} = \sqrt{ {GM_{\rm BH} \over r} - \eta v_{\rm turb}^{2} }~.
\end{equation}
One can infer from the above equation that 
\begin{equation}
M_{\rm enc} = M_{\rm BH} - \Delta M_{\rm turb}~,
\end{equation}
where $M_{\rm enc}$ $\equiv$ $rv_{\rm orb}^{2}/G$ is the enclosed mass measured from the observed orbital velocity of a maser spot and $\Delta M_{\rm turb}$ $\equiv$ $\eta$$rv_{\rm turb}^{2}$/G gives the mass underestimation caused by wind disturbance if one uses $M_{\rm enc}$ to estimate $M_{\rm BH}$.

In the 3rd column of Table \ref{table:mass_underestimation}, we list our conservative estimates of turbulent velocity dispersions $v_{\rm turb}$ for double-peaked megamasers in Table \ref{table:luminosities}. We estimate $v_{\rm turb}$ from the velocity scatter seen in the rotation curves of these double-peaked systems. NGC 5728 is not included here because this system is determined to be in a wind and shows a rising rotation curve. In the 4th column of the table, we show the ratio between $\Delta M_{\rm turb}$ and the BH mass estimate shown in Table \ref{table:luminosities}, with $\Delta M_{\rm turb}$ evaluated using the inner radius of the maser disk. One can see from this table that except for NGC 3079 and NGC 4945, $\Delta M_{\rm turb}$ is smaller than the BH mass error caused by uncertainty of the galaxy distance. Given the magnitude of $\Delta M_{\rm turb}$ shown in Table \ref{table:mass_underestimation}, one can infer that the effect of wind disturbance would not be large enough to account for the systematic difference of $M_{\rm BH}$ between the double-peaked and triple-peaked maser systems.

\section{Maser positions and velocities for NGC 5728, Mrk1, and IRAS
08452$-$0011 }

\begin{table*}
	\centering
	\caption{The Position and Velocity of each Maser Spot  in NGC
5728}
	\label{table:NGC5728data} 
	\begin{tabular}{rcccccc} % four columns, alignment for each
		\hline\hline
V$_{\rm op}$ & RA
& $\delta$RA  & Decl. 
& $\delta$DEC  & F$_\nu$  & $\sigma_F$ \\ 
(km~s$^{-1}$)  & (mas) 
& (mas)        & (mas)
& (mas) & (mJy/B) & (mJy/B) \\
	\hline
   209.01  &  $-$1.241  &     0.031  &     1.404  &     0.127  &    17.0  &     1.9   \\
   207.29  &  $-$1.291  &     0.031  &     1.612  &     0.123  &    15.9  &     1.7   \\
   205.57  &  $-$1.231  &     0.053  &     0.821  &     0.203  &    10.4  &     1.8   \\
   188.38  &  $-$1.134  &     0.034  &     0.166  &     0.122  &    18.5  &     2.0   \\
   186.66  &  $-$1.042  &     0.057  &  $-$0.026  &     0.172  &    13.1  &     1.8   \\
   119.58  &     1.202  &     0.033  &     0.262  &     0.137  &    13.6  &     1.8   \\
   117.86  &     1.196  &     0.053  &     0.481  &     0.216  &    10.3  &     1.7   \\
   116.14  &     1.158  &     0.051  &     0.234  &     0.162  &    10.8  &     1.8   \\
   114.42  &     1.162  &     0.042  &     0.036  &     0.159  &    11.9  &     1.8   \\
   -74.01  &     3.573  &     0.013  &     0.866  &     0.049  &    57.0  &     2.4   \\
  -170.18  &     4.669  &     0.028  &     2.232  &     0.127  &    17.3  &     2.0   \\
  -170.00  &     4.638  &     0.028  &     2.201  &     0.109  &    17.3  &     1.9   \\
  -195.74  &     5.682  &     0.032  &     3.502  &     0.115  &    16.7  &     1.9   \\
  -199.18  &     5.757  &     0.046  &     3.190  &     0.210  &    11.4  &     1.8   \\
                \hline     
\multicolumn{7}{|p{8 cm}|}{Note. Col.(1): Maser velocity relative to the systemic velocity of the galaxy ($V_{\rm sys}$ $=$ 2812 km~s$^{-1}$).
The velocity shown here is referenced to the LSR using
the optical definition;                                                
Col.(2)$-$Col.(5): East-west and north-south position offset and uncertainty
measured relative to the phase-reference center (RA $=$ 14:42:23.87213;
DEC $=$ $-$17:15:11.0165). The position uncertainty reflects measurement
errors only ; Col.(6)$-$Col.(7): Fitted peak intensity and its uncertainty
in mJy~beam$^{-1}$.   }\\                
	\end{tabular}

\end{table*}

\begin{table*}
	\centering
	\caption{The Position and Velocity of each Maser Spot  in Mrk 1}
	\label{table:Mrk1data} 
	\begin{tabular}{rcccccc} % four columns, alignment for each
		\hline\hline
V$_{\rm op}$ & RA
& $\delta$RA  & Decl. 
& $\delta$DEC  & F$_\nu$  & $\sigma_F$ \\ 
(km~s$^{-1}$)  & (mas) 
& (mas)        & (mas)
& (mas) & (mJy/B) & (mJy/B) \\
	\hline
   107.25  &  $-$0.157  &     0.032  &     0.351  &     0.067  &     5.4  &     0.8   \\
   103.77  &  $-$0.144  &     0.023  &     0.166  &     0.046  &     7.6  &     0.8   \\
   102.03  &     0.037  &     0.009  &     0.030  &     0.021  &    17.7  &     0.8   \\
   100.29  &     0.044  &     0.007  &  $-$0.004  &     0.016  &    23.0  &     0.8   \\
    98.54  &  $-$0.039  &     0.017  &     0.139  &     0.037  &    10.4  &     0.8   \\
    96.80  &  $-$0.012  &     0.048  &     0.253  &     0.101  &     4.3  &     0.8   \\
    95.06  &     0.110  &     0.035  &  $-$0.138  &     0.070  &     5.2  &     0.7   \\
    93.32  &     0.004  &     0.035  &     0.094  &     0.064  &     5.0  &     0.8   \\
    91.58  &     0.006  &     0.016  &     0.016  &     0.033  &    10.7  &     0.8   \\
    89.83  &     0.032  &     0.006  &     0.014  &     0.015  &    23.4  &     0.8   \\
    88.09  &     0.020  &     0.005  &     0.045  &     0.011  &    31.0  &     0.8   \\
    86.35  &     0.012  &     0.004  &     0.049  &     0.009  &    41.3  &     0.8   \\
    84.61  &     0.027  &     0.005  &     0.015  &     0.012  &    29.9  &     0.8   \\
    82.87  &     0.030  &     0.012  &     0.033  &     0.026  &    13.9  &     0.8   \\
    81.12  &     0.098  &     0.012  &  $-$0.063  &     0.024  &    15.1  &     0.8   \\
    79.38  &     0.137  &     0.010  &  $-$0.147  &     0.021  &    17.6  &     0.8   \\
    77.64  &     0.225  &     0.016  &  $-$0.216  &     0.031  &    11.6  &     0.8   \\
    75.90  &     0.254  &     0.015  &  $-$0.278  &     0.035  &    11.1  &     0.8   \\
    74.16  &     0.231  &     0.022  &  $-$0.242  &     0.056  &     7.0  &     0.8   \\
    72.42  &  $-$0.906  &     0.055  &     0.527  &     0.080  &     4.3  &     0.7   \\
    70.67  &  $-$0.175  &     0.035  &  $-$0.290  &     0.069  &     5.5  &     0.8   \\
    67.19  &  $-$0.254  &     0.040  &     0.470  &     0.088  &     4.7  &     0.8   \\
    65.45  &  $-$0.302  &     0.016  &     0.473  &     0.037  &    10.1  &     0.8   \\
    61.96  &  $-$0.964  &     0.050  &     0.622  &     0.081  &     4.0  &     0.8   \\
    60.22  &  $-$0.897  &     0.016  &     0.491  &     0.037  &     9.5  &     0.8   \\
    49.77  &  $-$1.430  &     0.040  &     0.422  &     0.094  &     4.2  &     0.8   \\
    44.55  &  $-$1.367  &     0.032  &     0.311  &     0.085  &     4.2  &     0.8   \\
    42.80  &  $-$1.380  &     0.031  &     0.161  &     0.097  &     4.8  &     0.8   \\
    35.84  &  $-$1.124  &     0.037  &  $-$1.162  &     0.078  &     4.5  &     0.7   \\
    34.09  &  $-$0.844  &     0.038  &  $-$1.121  &     0.083  &     4.0  &     0.7   \\
    32.35  &  $-$1.063  &     0.043  &  $-$1.192  &     0.124  &     3.6  &     0.7   \\
  -154.08  &  $-$0.347  &     0.031  &  $-$3.364  &     0.085  &     5.0  &     0.8   \\
  -155.82  &  $-$0.350  &     0.008  &  $-$3.298  &     0.018  &    19.8  &     0.8   \\
  -157.56  &  $-$0.343  &     0.022  &  $-$3.473  &     0.048  &     7.0  &     0.7   \\
  -159.30  &  $-$0.462  &     0.014  &  $-$3.481  &     0.037  &    10.2  &     0.8   \\
  -161.04  &  $-$0.445  &     0.021  &  $-$3.355  &     0.047  &     8.0  &     0.8   \\
  -162.78  &  $-$0.432  &     0.026  &  $-$3.414  &     0.066  &     5.9  &     0.8   \\
  -164.52  &  $-$0.335  &     0.039  &  $-$3.434  &     0.085  &     3.9  &     0.8   \\
  -166.26  &  $-$0.383  &     0.025  &  $-$3.383  &     0.069  &     5.3  &     0.8   \\
  -168.00  &  $-$0.317  &     0.037  &  $-$3.381  &     0.097  &     4.4  &     0.8   \\
  -169.73  &  $-$0.399  &     0.046  &  $-$3.325  &     0.112  &     4.0  &     0.7   \\
  -178.43  &  $-$0.271  &     0.037  &  $-$3.425  &     0.089  &     4.1  &     0.7   \\
  -180.17  &  $-$0.331  &     0.008  &  $-$3.484  &     0.019  &    20.1  &     0.8   \\
  -181.91  &  $-$0.345  &     0.012  &  $-$3.494  &     0.028  &    12.9  &     0.8   \\
  -192.34  &  $-$0.480  &     0.044  &  $-$2.862  &     0.074  &     4.5  &     0.7   \\
  -194.08  &  $-$0.547  &     0.019  &  $-$2.780  &     0.043  &     7.8  &     0.8   \\
  -195.82  &  $-$0.516  &     0.013  &  $-$2.798  &     0.031  &    10.7  &     0.8   \\
  -197.55  &  $-$0.542  &     0.018  &  $-$2.782  &     0.039  &     9.1  &     0.8   \\
  -199.29  &  $-$0.531  &     0.013  &  $-$2.784  &     0.033  &    10.6  &     0.8   \\
  -201.03  &  $-$0.549  &     0.020  &  $-$2.696  &     0.037  &     8.5  &     0.8   \\
  -202.77  &  $-$0.497  &     0.023  &  $-$2.754  &     0.058  &     5.2  &     0.7   \\
                 \hline     
\multicolumn{7}{|p{8 cm}|}{Note. Col.(1): Maser velocity relative to the systemic velocity of the galaxy ($V_{\rm sys}$ $=$ 4781 km~s$^{-1}$). The velocity shown here is 
referenced to the LSR using the optical definition;                                                
Col.(2)$-$Col.(5): East-west and north-south position offset and uncertainty
measured relative to the phase-reference center (RA $=$ 01:16:07.2093243; DEC $=$ $+$33:05:21.633601). The position uncertainty reflects measurement
errors only ; Col.(6)$-$Col.(7): Fitted peak intensity and its uncertainty
in mJy~beam$^{-1}$.   }\\                
	\end{tabular}

\end{table*}

\begin{table*}
	\centering
	\caption{The Position and Velocity of each Maser Spot  in IRAS
08452$-$0011}
	\label{table:J0847data} 
	\begin{tabular}{rcccccc} % four columns, alignment for each
		\hline\hline
V$_{\rm op}$ & RA
& $\delta$RA  & Decl. 
& $\delta$DEC  & F$_\nu$  & $\sigma_F$ \\ 
(km~s$^{-1}$)  & (mas) 
& (mas)        & (mas)
& (mas) & (mJy/B) & (mJy/B) \\
	\hline
   832.39  &  $-$0.547  &     0.085  &  $-$0.101  &     0.165  &    11.7  &     2.0   \\
   831.27  &  $-$0.537  &     0.066  &  $-$0.003  &     0.131  &    12.3  &     2.2   \\
   812.17  &  $-$0.499  &     0.066  &  $-$0.161  &     0.114  &    13.5  &     2.1   \\
   809.93  &  $-$0.462  &     0.044  &  $-$0.136  &     0.086  &    18.1  &     2.1   \\
   808.80  &  $-$0.460  &     0.034  &     0.004  &     0.091  &    22.6  &     2.2   \\
   807.68  &  $-$0.450  &     0.032  &     0.002  &     0.068  &    25.6  &     2.2   \\
   806.56  &  $-$0.488  &     0.042  &  $-$0.058  &     0.087  &    22.0  &     2.2   \\
   805.43  &  $-$0.495  &     0.059  &     0.097  &     0.103  &    17.1  &     2.1   \\
   804.31  &  $-$0.380  &     0.057  &     0.180  &     0.143  &    13.4  &     2.0   \\
   784.09  &  $-$0.578  &     0.065  &     0.127  &     0.139  &    14.6  &     2.0   \\
   782.97  &  $-$0.448  &     0.023  &  $-$0.085  &     0.051  &    33.8  &     2.2   \\
   781.84  &  $-$0.449  &     0.019  &  $-$0.113  &     0.047  &    41.8  &     2.2   \\
   780.72  &  $-$0.457  &     0.029  &  $-$0.111  &     0.064  &    31.1  &     2.3   \\
   779.60  &  $-$0.446  &     0.030  &     0.079  &     0.080  &    25.7  &     2.1   \\
   778.47  &  $-$0.462  &     0.037  &  $-$0.265  &     0.079  &    20.1  &     2.2   \\
   770.61  &  $-$0.447  &     0.042  &  $-$0.095  &     0.125  &    13.4  &     2.3   \\
   768.36  &  $-$0.532  &     0.072  &  $-$0.144  &     0.148  &    12.7  &     2.1   \\
   767.24  &  $-$0.408  &     0.058  &  $-$0.097  &     0.121  &    14.1  &     2.1   \\
   703.21  &  $-$0.551  &     0.114  &  $-$0.210  &     0.202  &    10.3  &     2.0   \\
   702.08  &  $-$0.279  &     0.116  &  $-$0.112  &     0.141  &    11.2  &     2.1   \\
    26.11  &  $-$0.438  &     0.082  &     0.113  &     0.155  &    12.3  &     2.3   \\
    24.99  &  $-$0.356  &     0.035  &     0.073  &     0.075  &    21.6  &     2.1   \\
    23.87  &  $-$0.403  &     0.049  &     0.113  &     0.106  &    18.2  &     2.0   \\
    22.76  &  $-$0.359  &     0.049  &     0.068  &     0.111  &    16.5  &     2.2   \\
    21.64  &  $-$0.291  &     0.050  &     0.142  &     0.126  &    18.4  &     2.1   \\
    20.52  &  $-$0.382  &     0.038  &     0.257  &     0.102  &    18.8  &     2.0   \\
    19.40  &  $-$0.390  &     0.061  &     0.233  &     0.119  &    15.1  &     2.2   \\
    18.29  &  $-$0.315  &     0.039  &  $-$0.075  &     0.091  &    18.9  &     2.2   \\
    17.17  &  $-$0.338  &     0.056  &  $-$0.069  &     0.122  &    17.3  &     2.1   \\
    16.05  &  $-$0.352  &     0.056  &     0.085  &     0.107  &    15.6  &     2.2   \\
    14.93  &  $-$0.387  &     0.063  &     0.107  &     0.134  &    14.5  &     2.3   \\
    13.81  &  $-$0.393  &     0.045  &  $-$0.029  &     0.095  &    20.0  &     2.3   \\
    12.70  &  $-$0.372  &     0.026  &     0.133  &     0.059  &    34.7  &     2.3   \\
    11.58  &  $-$0.363  &     0.032  &     0.032  &     0.072  &    27.0  &     2.1   \\
    10.46  &  $-$0.328  &     0.039  &  $-$0.024  &     0.091  &    20.8  &     2.1   \\
     9.34  &  $-$0.385  &     0.054  &  $-$0.027  &     0.125  &    14.2  &     2.2   \\
  -633.37  &  $-$0.142  &     0.044  &     0.414  &     0.135  &    13.1  &     2.0   \\
  -634.48  &  $-$0.083  &     0.057  &     0.202  &     0.168  &    11.8  &     2.1   \\
  -656.74  &  $-$0.281  &     0.065  &     0.307  &     0.117  &    11.9  &     2.0   \\
  -666.76  &  $-$0.142  &     0.087  &     0.208  &     0.184  &    10.3  &     2.1   \\
  -670.10  &  $-$0.186  &     0.046  &     0.452  &     0.106  &    15.6  &     2.2   \\
  -671.21  &  $-$0.173  &     0.049  &     0.324  &     0.126  &    14.8  &     2.1   \\
  -673.43  &  $-$0.186  &     0.055  &     0.309  &     0.122  &    15.2  &     2.2   \\
  -674.55  &  $-$0.231  &     0.049  &     0.339  &     0.104  &    16.8  &     2.0   \\
  -739.10  &  $-$0.213  &     0.055  &     0.167  &     0.111  &    15.5  &     2.1   \\
  -740.21  &  $-$0.330  &     0.062  &     0.492  &     0.141  &    11.5  &     2.1   \\
                 \hline     
\multicolumn{7}{|p{8 cm}|}{Note. Col.(1):  Maser velocity relative to the systemic velocity of the galaxy ($V_{\rm sys}$ $=$ 15262 km~s$^{-1}$). The velocity shown here is referenced to the LSR using the optical definition;                                                
Col.(2)$-$Col.(5): East-west and north-south position offset and uncertainty
measured relative to the phase-reference center (RA $=$ 08:47:47.70278; DEC $=$ $-$00:22:51.01505). The position uncertainty reflects measurement
errors only ; Col.(6)$-$Col.(7): Fitted peak intensity and its uncertainty
in mJy~beam$^{-1}$.   }\\                
	\end{tabular}

\end{table*}

{\bf Data Availability : }The data underlying this article are available in the article and in its online supplementary material.

%%%%%%%%%%%%%%%%%%%%%%%%%%%%%%%%%%%%%%%%%%%%%%%%%%



\bibliographystyle{mnras}

%\begin{thebibliography}{99}
%\end{thebibliography}

%%%%%%%%%%%%%%%%%%%%%%%%%%%%%%%%%%%%%%%%%%%%%%%%%%

%%%%%%%%%%%%%%%%% APPENDICES %%%%%%%%%%%%%%%%%%%%%

%%%%%%%%%%%%%%%%%%%%%%%%%%%%%%%%%%%%%%%%%%%%%%%%%%

% Don't change these lines
\bsp	% typesetting comment
\label{lastpage}
\end{document}